\documentclass[11pt,a4paper]{article}
\usepackage{ifpdf} \usepackage{ae} \usepackage[T1]{fontenc}
\usepackage[ansinew]{inputenc} \usepackage{mathrsfs}
\usepackage{amsmath} \usepackage{amssymb} \pdfoutput=1
\usepackage{amsmath,amssymb,amsfonts,a4wide,graphicx,bm,times,psfrag,wrapfig,sidecap}
\usepackage{cite} \usepackage[colorlinks=true,linkcolor=black,
citecolor=black, urlcolor=black]{hyperref}
\numberwithin{equation}{section} \makeatletter
\let\old@startsection=\@startsection \renewcommand{\@startsection}[6]
{\old@startsection{#1}{#2}{#3}{#4}{#5}{#6\mathversion{bold}}}
\makeatother \def\O{\Omega} 
\def\defeq{\stackrel{\text{def}}=} \newcommand\re[1]{({\ref{#1}})}
\def\be{\begin{eqnarray} } \def\ee{\end{eqnarray}} 
\def\IR{{\mathbb{R}}} \def\IZ{{\mathbb{Z}}} \def\IC{{\mathbb{C}}}
 \def\IC{{\mathbb{C}}}
\def\no{\nonumber} \def\la{\label} \def\l {\lambda} \def\CA{{\cal A}}
\def\CD{{\cal D}}   \def\CB{{\cal
B}} \def\CO{{ \mathcal{ O} }}   \def\o{\omega} \def\CC{ {\mathcal C}}
 \def\Tr{{\rm Tr}} \def\Li{ \text{Li}_2}
\def\d{\delta} \newcommand{\caS}{{\mathscr S}}
\newcommand{\caA}{{\mathscr A}} \newcommand{\caY}{{\mathscr Y}}
\newcommand{\caZ}{{\mathscr Z}} \newcommand\spinup[1]{\!\!
\uparrow\uparrow.  ..^{\!\!\!  \!\!  #1}\uparrow \!}
\newcommand\spindown[1]{\!\!\downarrow\downarrow.  ..^{\!\!\!  \!\!
#1}\downarrow \!  } \newcommand\braup[1]{\, \langle \uparrow ^{
#1}\!\!  |} \newcommand\ketup[1]{| \!  \uparrow ^{ #1} \rangle}
\newcommand\bradown[1]{\, \langle \downarrow^{#1}\!\!  | }
\def\slp{\slash\!\!\!  p } \def\ket{ | 0 \rangle} \def\bra{ \langle 0
| } \def\bS{{\mathbb S}} \def\({\left(} \def\){\right)}
\def\<{\left\langle\,} \def\>{\, \right\rangle} \def\[{\left[}
\def\]{\right]}  \def\hf{ {\textstyle{1\over 2}} }
\def\hfi{ {\textstyle{i\over 2}} } \def\CA{{\cal A}} 
\def\CB{{\cal B}} \def\CO{{ \mathcal{ O} }} 
 \def\CK{{ \mathcal{ K} }}  \def\o{\omega} \def\CC{ {\mathcal C}}
\def\CT{ {\mathcal T}} \def\CN{{ \cal N}}     
\def\CY{{\cal Y}} \def\p{\partial} \def\a{\alpha} 
  \def\s{\sigma} 
 \def\th{\theta}  
\def\Tr{{\rm Tr}} \def\Li{ \text{Li}_2} \def\d{\delta}
\def\sla{\slash\!\!  \!} \def\zz{{ { \mathbf z} }} \def\uu{ { {\bf u}
}} \def\vv{ { \bf v}} \def\ww{ { \bf w }} \def\thth{ { \bm {\th } }}
\def\vv { { \bf v }}  \def\k{\kappa}  \def\llangle{ \langle\!  \!\langle } \def\rrangle{
\rangle\!\!\rangle } \def\vvert{ |\!\!  | }
\def\az{{_{(1)}}}
\def\za{{_{(2)}} }  
  \def\bC{{\mathbb  C}}
      \def\bB{{\mathbb  B}}

    \def\bS{{\mathbb  S}}

\begin{document}

\thispagestyle{empty}

\begin{flushright}
  IPhT/t12/035  
\end{flushright}

\vspace{1cm}
\setcounter{footnote}{0}

\begin{center}

{\Large\bf  Three-point function of semiclassical states at weak coupling   }

\vspace{20mm} 

Ivan Kostov\footnote{\it Associate member of the Institute for Nuclear
Research and Nuclear Energy, Bulgarian Academy of Sciences, 72
Tsarigradsko Chauss\'ee, 1784 Sofia, Bulgaria} \\[7mm]
 
{\it Institut de Physique Th\'eorique, CNRS-URA 2306 \\
	     C.E.A.-Saclay, \\
	     F-91191 Gif-sur-Yvette, France} \\[5mm]

\end{center}

\vskip9mm

\vskip18mm

\noindent{ We give the derivation of the previously announced analytic
expression for the correlation function of three heavy non-BPS
operators in $\CN=4$ super-Yang-Mills theory at weak coupling.  The
three operators belong to three different $su(2)$ sectors and are dual
to three classical strings moving on the sphere.  Our computation is
based on the reformulation of the problem in terms of the Bethe Ansatz
for periodic XXX spin-1/2 chains.  In these terms the three operators
are described by long-wave-length excitations over the ferromagnetic
vacuum, for which the number of the overturned spins is a finite
fraction of the length of the chain, and the classical limit is known
as the Sutherland limit.  Technically our main result is a factorized
operator expression for the scalar product of two Bethe states.  The
derivation is based on a fermionic representation of Slavnov's
determinant formula, and a subsequent bosonisation.  }

\newpage
\setcounter{footnote}{0}
%

\section{Introduction}
\label{sec:Intro}

 During the last decade, starting with the pioneer paper by Minahan
 and Zarembo \cite{Minahan:2002ve}, a vast integrable structure has
 been unveiled in the $\mathcal{N}=4$ supersymmetric Yang-Mills (SYM)
 theory and in its dual supersymmetric string theory in the
 $AdS_5\times S^5$ space-time.  The integrable structure, known by
 the name of  Integrability, was formulated in terms of an effective long
 range spin chain having $PSL(2,2|4)$ symmetry.  For a recent review
 see \cite{Beisert-Rev}.

The spectral problem, i.e. the problem of computing the anomalous
dimensions of the gauge-invariant states in SYM and their string
counterparts, is nowadays considered as conceptually solved.
Spectacular progress was achieved in the last years in the computation
of the gluon amplitudes \cite{AGM-Bubble, AMSV} and Wilson loops
\cite{2012arXiv1203.1617D, 2012arXiv1203.1913C}.  The next step
towards the complete solution is to compute the three-point function
of gauge-invariant operators representing traces of products of
fundamental fields.  This is obviously an extremely hard problem, but
the encouraging developments over the last several years
\cite{2004JHEP...08..055O,2004JHEP...09..032R,EGSV,
Escobedo:2011xw,GSV} raise the hope that Integrability can be used for
this problem as well.

Of special interest are the correlation functions of one-trace
operators in the classical limit when the length of the traces is very
large.  Such operators are dual to extended classical strings in the
AdS$_5\times$S$^5$ background.  This correspondence allows to approach
the problem both at strong and at weak coupling.  The classical, or
long-trace, operators are described in terms of algebraic curves,
which appear as finite-gap solutions of the Bethe equations in the
classical limit, or as classical solutions of the string sigma model
\cite{Kazakov:2004qf, Kazakov:2004nh, Beisert:2005bm} (see also the
review \cite{Schafer-Nameki:2010aa}).

On the string theory side, the problem of computing the correlation
function of operators dual to classical spinning string solutions was
addressed by several authors \cite{Janik:2010aa,
Zarembo:2010ab,Buchbinder:2010ae,2011arXiv1109.6262J,
2012PhRvD..85b6001B, 2011arXiv1106.0495K, 2011arXiv1110.3949K}.  The
three-point function of such operators can be thought of as a
classical tunneling amplitude.  To compute it, one should find the
appropriate Euclidean classical solution for a world sheet embedded in
$AdS_5\times S^5$ and having the topology of a sphere with three
punctures.  However, there is not yet a consensus among the active
workers on the field about the criteria to distinguish the relevant
classical solution, neither there is an unambiguous prescription about
how to construct the vertex operators associated with the punctures.
The only case when the complete answer is known is that of two heavy
and one light operators
\cite{2010JHEP...11..141C,Zarembo:2010ab,Roiban:2010aa}.

Alternatively, one can start with three one-trace operators in the
weakly coupled gauge theory, where the computation of the correlation
functions is a well defined problem.  At tree level it is sufficient
to count all possible planar sets of Wick contractions between the
three operators.  It was pointed out by Roiban and Volovich
\cite{2004JHEP...09..032R} that the calculation of correlation
functions of gauge invariant operators reduces at the level of the
spin chain to the calculation of the scalar products of states
constructed out of B and C operators in the Algebraic Bethe Ansatz
\cite{Faddeev:1979gh}.  A systematic study of the case when the three
operators belong to three different $su(2)$ sectors of the theory was
presented by Escobedo, Gromov, Sever and Vieira \cite{EGSV}.  The
tree-level correlation function of three $su(2)$ operators was
expressed in \cite{EGSV} in terms of scalar products of Bethe states
in a periodic XXX spin chain with spin 1/2.  The formalism developed
in \cite{EGSV} was later applied to compute the correlation function
of two heavy and one (or more) light operators
\cite{Escobedo:2011xw,2011arXiv1107.5580C}, and the comparison with
the strong coupling results on the string theory side
\cite{2010JHEP...11..141C,Zarembo:2010ab,Roiban:2010aa} showed a
precise match.  The limit of three heavy, or classical, operators was
then obtained in \cite{GSV} in the case when one of the operators is
BPS type, that is, protected by the supersymmetry.  The main result of
\cite{GSV} is an elegant analytic formula in the form of a contour
integral.  The derivation was based on the expansion formula for the
scalar products of two generic Bethe states due to Korepin
\cite{korepin-DWBC}.

In this paper we tackle the general case when neither of the three
heavy operators is BPS type.  We compute, at tree-level, the
correlation function of three non-protected classical operators
belonging to three different $su(2)$ sectors of the gauge theory.  The
principal object to be computed is the restricted scalar product of
two Bethe states, in which part of the rapidities are frozen to a
special value.  Our result, which is a generalization of the main
result of \cite{GSV}, was announced in the short note \cite{3pf-prl}.
The analytic expression found in \cite{3pf-prl} is based on a
factorization formula, which follows from the representation of the
structure constant in terms of Slavnov-like determinants
\cite{NSlavnov1}, proposed recently by Foda \cite{Omar,
2012arXiv1203.5621F, 2012arXiv1202.6553A}.

 The plan of the paper is as follows.  In Section \ref{sec:XXX} we
 review the basic notions about the XXX spin chain we are going to
 use, including the Gaudin norm, the Slavnov determinant formula for
 the scalar product, and its restricted version.  Following Foda
 \cite{Omar}, we will perform the computations for the inhomogeneous
 spin chain, characterized by a set of external parameters
 (impurities) associated with the sites of the chain.  Such a
 deformation of the problem allows to avoid ambiguous expressions
 containing poles and zeroes on the top of each other. 
  In Section \ref{sec:fermionicSL} we derive a
 representation of the Slavnov determinant in terms of free chiral
 fermions, and then perform bosonization.  As a side result, we obtain
 a novel and potentially useful representation of the Izergin
 determinant for the domain wall partition function (DWPF) of the
 six-vertex model.  In the limit when the number of Bethe roots tends
 to infinity, the bosonized expression decomposes into two computable
 factors, for which we find the analytic expression in Section
 \ref{sec:classical}.  In section \ref{sec:3pf} we derive the
 structure constant of three classical non-BPS operators in the
 $sl(2)$ sector of $\CN=4$ SYM.

 \section{Inner product of Bethe states in the inhomogeneous XXX
 chain}
\la{sec:XXX}

 \subsection{ The monodromy matrix}

The local fluctuation variable in a XXX spin 1/2 chain can be in two
states, $\downarrow$ and $\uparrow$, which can be thought of as a
basis of a two-dimensional linear space $V$ on $\IC$.  The spin chain
is characterized by an isotropic Hamiltonian
   \be H_{_{\text{XXX}}} =- \sum_{m=1}^L \( \s^+_m \s^-_{m+1} + \s^-_m
   \s^+_{m+1} + \hf \s^x_m \s^x_{m+1} \) .  \ee
 We will assume twisted periodic boundary conditions,
   \be \s_{m+L}^{\pm} = \k^\pm \, \s_m^\pm\, , \ee 
   which do not spoil the integrability, but allow to have a better
   control over the singularities.  If the twist \be \la{twist} \k
   \equiv {\k_- /\k_+} = e^{i\phi} \ee
is a pure phase, the twisted boundary conditions can be thought of as
the effect of turning on a magnetic flux of strength $ \phi$
\cite{PhysRevLett.7.46, PhysRevLett.74.816}.  For us $\k$ will be an
unrestricted complex parameter.
 
In the framework of the Algebraic Bethe Ansatz \cite{Faddeev:1979gh},
the spin chain is characterized by an $R$-matrix $R_{12}(u,v)$ acting
in the tensor product $V_1\otimes V_2$ of two copies of the target
space,
  \be \la{defR} R_{12}(u,v) = u-v + {iP_{12} }\, , \ee
where $P_{12} $ is the permutation operator \cite{Fad-Tacht-84}.  We
will consider the {\it inhomogeneous } XXX spin chain, characterized
by background parameters (impurities) $\thth=\{ \th_1,\dots , \th_L\}$
associated with the $L$ sites of the chain.  The twisted monodromy
matrix $\CT_a (u)\in \text{End}(V_a)$ is defined as the product of the
$R$-matrices along the spin chain and a twist matrix $K= \(^{\k_+\
0}_{\, 0 \ \ \, \k_-}\)$,
     \be \la{defTa} \CT_a(u) \equiv K R_{1 a}(u, \th_1 + \hf i)\,
     R_{2a}(u, \th_2+\hf i)\dots R_{La}(u, \th_L+\hf i) =K
     \begin{pmatrix} {\CA}(u) &{ \CB}(u) \\
    \CC(u)  &  \CD(u)
    \end{pmatrix} .
 \ee
 For the homogeneous XXX spin chain all $\th_m $ are equal to 0.  The
 advantage of introducing the twist and the inhomogeneity parameters
 is that the expressions for some scalar products, which are ambiguous
 for $\th_m =0$ and $\k=1$, becomes well defined for generic $\th_m $
 and $\k$.

The matrix elements $\CA,\CB,\CC, \CD$ are operators in the Hilbert
space $V= V_1\otimes\dots V_L$ of the spin chain.  The commutation
relations between the elements of the monodromy matrix are determined
by the relation
  \be\la{RTT} R_{12}(u-v) \CT_1(u) \CT_2(v) = \CT_2(v) \CT_1(u) \,
  R_{12}(u-v) \ee
 which follows from the Yang-Baxter equation for $R$.  As a
 consequence of \re{RTT}, the  operators $\CB(u)$, $\CC(u)$,
 and the transfer matrices
  \be \la{defTrM} \CT (u) \equiv\Tr_a [ \CT_a(u)] = \CA(u) +\CD(u),
   \ee
  form commuting families:
  \be \la{CCR} [\CB(u), \CB(v)]=[\CC(u),\CC(v)]=[\CT(u), \CT(v)]=0
  \quad \text{for}\ \  u, v\in\IC.
  \ee

    \subsection{ The Hilbert space as a Fock space for the
    pseudo-particles (the magnons)}

One can give the Hilbert space $V$ a structure of a Fock space
generated by the action of the operators $\CB(u)$ on the pseudo-vacuum
$ \braup{L}= \big|\, \spinup{L } \big\rangle $.  The pseudo-vacuum is
an eigenvector for the diagonal elements $\CA$ and $\CD$ and is
annihilated by $\CC$:
 \be \la{acket} \CA(u)\ \ketup{L}= a(u)\ \ketup{L}, \ \ \
 \CD(u)\ketup{L} = d(u)\ketup{L}, \ \ \ \CC(u)\ketup{L}= 0.  \ee
 The dual Bethe states are generated by the action of the
 $\CC$-operators on the dual pseudo-vacuum $\braup{L}= \big\langle
 \spinup{L } \big|$, which has the properties
 \be \la{acbra} \braup{L}\ \CA(u) = a(u)\braup{L} , \ \ \ \bra\CD(u) =
 d(u) \braup{L}, \ \ \ \braup{L} \CB(u) = 0.  \ee
The functions $a(u)$ and $d(u)$ depend on the representation of the
algebra \re{RTT}, while the $R$-matrix \re{defR} is universal.  For
the inhomogeneous XXX spin $1/2$ magnet the functions $a(u)$ and
$d(u)$ are given, according to \re{defTa}, by
 \be
 \la{defad}
 \begin{aligned}
 a(u) &=\k_+ \prod_{m=1}^L \(u- \th_m+i/2\) 
 ,
 \\
 d(u) &= \k_-\prod_{m=1}^L\( u- \th_m -i/2\) 
 .
\end{aligned}
 \ee

The algebraic construction of the Hilbert space does not use the
particular form of the functions $a(u) $ and $d(u)$, which can be
considered as free functional parameters
\cite{korepin-DWBC}.\footnote{In this case one speaks of generalized
$SU(2)$ model.} The space of states is a closure of the linear span of
all vectors of the form
\be \la{genstateket} \big| \uu \big\rangle = \prod_{j=1}^{N} \CB(u_j )\ 
\ketup{L} .  \ee
Here and below we will use the shorthand notation \footnote{We will
indicate the cardinality $N$ only if this is required by the
context.}
 \be \la{defsetrap} \uu = \uu_{_N} = \{u_1,\dots u_{_N}\}.  \ee
The operator $\CB(u)$ can be viewed as a creation operator of a
pseudo-particle (magnon) with momentum $p= \log{u+i/2\over u-i/2}$.
Similarly, the dual space of states is a closure of the linear span of
all vectors of the form\footnote{ With  the convention
 $\CB(u)^\dag = - \CC(u^*)$ , the 
state dual to $|\vv\rangle $ is $(-1)^N \langle \vv^*|$. 
}
 \be \la{genstatebra} 
 \big\langle \uu \big|= \braup{L}\, \prod_{j =1}^N
 \CC(u_j) .  \ee
Such states are   generic, or off-shell, Bethe states.
A Bethe state \re{genstateket} which is also an eigenstate of the
transfer matrix \re{defTrM} is called  on-shell state.  For a chain
of length $L$ there are $2^L$ linearly independent on-shell states.
Applying repeatedly the RTT relations \re{RTT} for the upper
triangular elements of the monodromy matrix, namely
   \be
   \begin{aligned}
   \CA(v) \CB(u) &= \textstyle{u-v+i\over u-v} \ \CB(u) \CA(v)
   -\textstyle{i\over u-v} \CB(v) \CA(u), \\
   \CD(v) \CB(u) &= \textstyle{u-v-i\over u-v} \ \CB(u) \CA(v) 
   +  \textstyle{i\over u-v} \CB(v) \CD(u),
   \end{aligned}
   \ee
one obtains \cite{Fad-Tacht-84} the on-shell condition
 for the rapidities $\uu= \{u_1,\dots
u_N\}$, which coincide with the (twisted) Bethe equations
\be \la{BAEXXX} {d(u_j )\over a(u_j )} \prod_{k =1}^N {u_j - u_k
+i\over u_j -u_k - i} =-1, \quad a=1, \dots, N. \ee
 Since the XXX Hamiltonian is
hermitian, the set of roots $\uu$ of a Bethe eigenstate must be
invariant under complex conjugation.  The corresponding eigenvalue
$T_\uu (u)$ of the transfer matrix is
    \be \la{eigenvalueT} T_{\uu}(u) = a(u) {Q_{\uu}(u-i)\over
    Q_{\uu}(u)} + d(u) {Q_{\uu}(u+i)\over Q_{\uu}(u)} , \ee
where $Q_\uu(u)$ is the Baxter polynomial 
\be \la{defQ} Q_\uu(u) \equiv \prod_{j =1}^N (u- u_j ) .  \ee

The Bethe equations \re{BAEXXX} can be considered as saddle-point
equations for the Yang-Yang functional \cite{YangYang-1966,
Faddeev:1979gh}, which we denote by $\caY_\uu$,
\be \p_{u_j } \caY_{_\uu} = 2\pi i n_j \quad (a=1,\dots, N; \ n_j \in
\IZ).  \ee
For the twisted periodic XXX$_{1/2}$ spin chain,  the Yang-Yang
functional is given by
\be \caY_{\uu}&=& \sum_{j =1}^N \sum_{m=1}^L (u_j -\th_m+\hf i) \log
(u_j -\th_m +\hf i)- (u_j -\th_m-\hf i) \log (u_j -\th_m -\hf i) \no
\\
& +& \sum_{j <k} \[(u_j -u_k +i )\log( u_j -u_k +i)- (u_j -u_k -i)
\log( u_j -u_k -i)\]
      \\
& +& i\phi  \ \sum_{j =1}^N u_j .  \ee

 \subsection{The pseudo-momentum   }
     
For each set of points $\uu= \{u_j \}_{j =1}^N $, define the function
$p_\uu(u)$, called {\it pseudomomentum}\footnote{There is no complete
consensus about the terminology.  The quantity we refer to as
pseudo-momentum is related to the {\it counting function} $Z(u)$ by
$e^{2i p_\uu(u)} = (-1)^{N+L} e^{ - i Z(u)}$.}
\be \la{defquazimomentum} e^{2i p_\uu (u) } \equiv {d(u)\over a(u)}\
\, {Q_\uu(u+i)\over Q_\uu(u-i)} = {\k}\,
{Q_\thth(u-i/2)\over Q_\thth(u+i/2)} \, {Q_\uu(u+i)\over Q_\uu(u-i)} .
\ee
 Here and below we will  use the notation  \re{defQ} for 
  products over rapidities. 
In terms of the pseudo-momentum, the Bethe equations \re{BAEXXX} read
 \be \la{BAE} e^{2ip_\uu(u)}=-1, \quad u\in \uu.  \ee
The pseudo-momentum is determined by \re{defquazimomentum} modulo
$i\pi$.  To characterize this function completely, it is necessary to
specify a set of integers (mode numbers) $\{ n_j\}_{j=1}^N$, not
necessarily different, so that $p(u_j) = \pi n_j$.

\subsection{Slavnov's determinant formula for the inner  product}

The scalar product of two generic Bethe states can be computed from
the algebra \re{RTT}, the relations \re{acket} and \re{acbra}, and the
convention $\langle \uparrow^L|\uparrow^L\rangle=1$.  In general, the
scalar product $\langle \vv|\uu\rangle$ of two Bethe states is given
by a double sum over partitions of the sets $\uu=\uu_{_N}$ and
$\vv=\vv_{_N}$, which becomes increasingly difficult to tackle when
number of magnons $N$ becomes large.  A significant simplification
occurs when one of the two sets of rapidities, say $\uu$, satisfies
the Bethe equations \re{BAEXXX}.  It was discovered by N. Slavnov
\cite{NSlavnov1} that in this case the scalar product is a
determinant.  This is true for all integrable models with $A_1^{(1)}$
type R-matrix.
    
Let the set $\uu$ satisfy the Bethe equations \re{BAEXXX}.  Then
$\vert \uu\rangle $ is an eigenvector for the transfer matrix with
eigenvalue $T_\uu(u)$, given by eq.  \re{eigenvalueT}.  It was shown
by Slavnov \cite{NSlavnov1,nikitaslavnov} that the scalar product with
a generic Bethe state $\langle \vv |$ is proportional to the
determinant of the matrix of the derivatives of $T_\uu(u)$, evaluated
at the points of $\vv$,
   \be \la{defSuv0}
    \langle \vv |\uu\rangle &=&\prod_{j =1}^N a(v_j )
   d(u_j )\ \caS_{\uu, \vv}\, ; \ee
   \be \caS _{\uu, \vv} &\defeq& {1\over \prod_{j =1}^N a(v_j )}\ {
   \det_{jk} {\p\over \p u_j } T_{\!  _\uu}\!  (v_k ) \over \det_{jk}
   {1\over u_j -v_k } } .  \la{defSuv} \ee
 (Eqs.  \re{defSuv0}-\re{defSuv} are equivalent to eq.  (6.16) of
 \cite{nikitaslavnov}.)  The explicit expression for the matrix of the
 derivatives $\p_{u_j }T_\uu(v_k )$ is
    \be\la{defOmo} -\p T_{_\uu}(v_k )/ \p u_j &=& t(u_j -v_k ) \,
    a(v_k ) {Q_{_\uu} (v_k -i)\over Q_{_\uu} (v_k )} - t(v_k -u_j ) \,
    d(v_k ) {Q_{_\uu} (v_k +i)\over Q_{_\uu} (v_k )}\, \no \\
    &=& a(v_k)  {Q_{\uu} (v_k -i)\over Q_{\uu} (v_k
    )} \ 
    \O(u_j, v_k)
     , \ee
where the kernel $\O(u,v)$ is defined by
  \be \la{defOhat} \O(u,v)= t(u-v) - e^{2 i p_\uu(v)} \ t(v-u)\, ,
  \qquad t(u) = {1\over u } - {1\over u+i}\, .  \ee
 One can simplify the Slavnov determinantt \re{defSuv} by noticing
 that $ \prod _{k=1}^N {Q_\uu(v_k-i)\over Q_\uu(v_k)} =
 {\det_{jk}{(u_j-v_k)^{-1}} \over \det_{jk}{(u_j-v_k+i)^{-1}}}.  $
 The resulting expression is
 \be \caS_{\uu, \vv}
 &=& {\det_{j k}\O(u_j , v_k ) \over \det_{j k} {1\over u_j -v_k
 +i} }  \la{SLhat}.
  \ee

In order to get  rid of the factors $a(v_j )\, d(u_j )$ in \re{defSuv0},
we  rescale  the annihilation/creation operators as
 \be
  \CC(u)\to \bC(u) = { \CC(u)\over a(u)}, \quad
   \la{SLNN} \CB(u) \to \bB(u) = {\CB(u)\over d(u)}  \ee
and denote the rescaled inner product by $\llangle \uu\vvert
\vv\rrangle$:
     \be \la{thicksp} \llangle \vv \vvert \uu\rrangle \defeq \langle
     \uparrow^{L}\vert \prod_{v\in\vv} \bC(u) \ \prod_{u\in\uu} \bB(u)
     \vert \uparrow^{L} \rangle = \caS _{\uu, \vv}\, .  \ee
  Depending on the context, sometimes we will indicate the set 
  of the inhomogeneity parameters and/or the cardinalities of the 
  sets $\uu,\vv$ and $\thth$,
  \be
     \llangle \vv \vvert \uu\rrangle 
     =   \llangle \vv \vvert \uu\rrangle _{\thth}
     =   \llangle \vv_N \vvert \uu_N\rrangle_{\thth_L} .
     \ee

      \subsection{The Gaudin norm}

 The hermitian conjugation compatible with the scalar product
 \re{defSuv0} is $\CC(u) = - \CB(\bar u)^\dag$,
 or 
 \be
 \bC(u) = {d(u)\over a(u)}\ \bB(\bar u)^\dag.
 \ee
   Assuming that the
 sets $\uu$ and $\vv$ are invariant under complex conjugation, $\bar
 \uu =\uu, \, \bar\vv =\vv$, and taking the limit $\vv\to\uu$ in
 \re{SLhat}, one reproduces the determinant expression for the square
 of the norm of a Bethe eigenstate conjectured by Gaudin
 \cite{Gaudin-livre,PhysRevD.23.417} and proved by Korepin in
 \cite{korepin-DWBC}.  The square of the norm is proportional to the
 Hessian of the Yang-Yang functional.  In our normalization
\be \la{Gaudinnorm} \llangle \uu\vvert\uu\rrangle &=&
{\det _{j k} \(i {\p^2 \CY_{_\uu}\over \p u_j \p u_k } \) \over
\det_{jk}{1\over 1+i(u_j -u_k )}} .  \ee
 The explicit expression for the matrix of the second derivatives is
 \be \la{ddCY} i {\p^2 \CY_{_\uu}\over \p u_j \p u_k }
 &=& {2\over (u_j -u_k )^2+1} - \d_{j _k } \( \sum_{l=1}^N {2\over
 (u_j -u_l )^2+1} - \sum_{j=1}^L {1\over (u_j -\th_m )^2+{1\over 4}}
 \).  \ee

\section{ Operator factorization formula for the inner product}
 \la{sec:fermionicSL}

 \subsection{Slavnov's determinant as a fermionic Fock space
 expectation value}
 \label{ssec:fr1}

The Slavnov determinant expression \re{SLhat} for the scalar product
can be formulated in terms of free fermions and represents a
tau-function of the KP/Toda hierarchy \cite{JPSJ.62.1887,FWZ,FoSh}.
For the Izergin determinant this was shown and used in \cite{
PZinn-review6v, CPZinn-2010, 2012arXiv1203.5621F}, see also the review
paper \cite{2010arXiv1003.3071T}.  The two sets of Toda times are
related to the moments of the two sets of rapidities,
$\uu=\{u_1,\dots, u_N\} $ and $\vv=\{v_1,\dots, v_N\}$.

The fermion representation we are going to use here is not the most
natural one from the point of view of integrable hierarchies, but it
reveals a hidden factorization property of the scalar product, which
can be used to find an analytic expression in the thermodynamical
limit $N, L\to\infty$.

Introduce a chiral Neveu-Schwarz fermion living in the rapidity
complex plane and having mode expansion
\be \la{pzpo} \psi (u)= \sum_{r\in \IZ+ {1\over 2}}\psi_{r}\ u^{-r-
{1\over 2}}, \ \ \ \ \bar \psi (u)= \sum_{r\in \IZ+ {1\over 2}} \bar
\psi_{ r}\ u^{r- {1\over 2}} .  \ee
The fermion modes are assumed to satisfy the anticommutation relations
\be\la{cpmto} [\bar \psi _{ r}, \bar \psi _{ r'} ]_+= [ \psi _{ r},
\psi _{ r'} ]_+= 0\, , \quad [ \psi _{ r}, \bar \psi _{ r'} ]_+=
\delta_{r, r'}\, ,  \ee
and the left/right vacuum states are defined by
\be\la{mnfio2} \bra \psi_{-r}= \bra \bar \psi_{r} = 0\ \ \text{and}\ \
\ \psi_{r}\, \ket =\bar \psi_{-r} \ket = 0,\ \ \ \text{for} \ r> 0 .
\ee
The operator $\bar \psi_r$ creates a particle (or annihilates a hole)
with mode number $r$ and the operator $\psi_r$ annihilates a particle
(or creates a hole) with mode number $r$.  The particles carry charge
1, while the holes carry charge $-1$.  The charge zero vacuum states
\re{mnfio2} are obtained by filling the Dirac see up to level zero.
The left vacuum states with integer charge $\pm N$ are constructed as
 \be
 \la{defchargeN}
 \langle N|\ = \ 
 \begin{cases}
   \bra \psi_{1\over 2} \dots \psi_{N-{1\over 2}} & \text{ if}\ N>0 ,
   \\
  \bra \bar \psi_{-{1\over 2}}\dots \bar\psi_{-N+{1\over 2}} &
  \text{if} \ N<0 .
\end{cases}
   \ee
 Any correlation function of the operators \re{pzpo} is a determinant
 of two-point correlators
\be \la{opepsi} \bra \psi(u) \bar \psi(v)\ket = {1\over u-v}\, .  \ee

Obviously, the Slavnov kernel  \re{defOhat} can be
represented as the correlation function of two fermionic operators,
located at the points $u$ and $v$ of the rapidity plane,
\be \la{SlM1} \O(u,v)&=& \bra \[
\bar \psi(v)  -  e^{2i p_{_\uu}(v)} \,\bar\psi(v+i) \]
 \[ \psi(u) -  \psi(u+i) \]\ket.
\ee
 The determinant of the matrix $\O(u_j, v_k)$ is equal to the
 correlation function of $N$ pairs of such operators, and the Slavnov
 inner product \re{SLhat} can be given the following Fock space
 representation,
\be \caS _{\uu, \vv}
&=& {\bra \prod_{j =1}^N \[ \psi(u_j ) - \psi(u_j +i) \]
 \ \prod_{k =1}^N \[ \bar\psi(v_k ) -e^{2i p_{_\uu}(v_k )}\, \bar
 \psi(v_k +i) \]
\ket \over \bra \prod_{j =1}^N \psi(u_j +i ) \prod_{k =1}^N\bar
\psi(v_k )\ket } \no .  \no\\
\la{fermrepb}
\ee

Our aim is to rewrite \re{fermrepb} in a form convenient for taking
the limit $N\to\infty$.  For that we first we transform the
denominator, using the Cauchy identity, to
\be \la{Cauchydet} \CK_{\uu,\vv}&=& \bra \prod_{j =1}^N \psi(u_j
+i)\prod_{k =1}^N\bar \psi(v_k ) \ket \no \\
&=&  \det_{jk} {1\over u_j  -v_k +i }
\no
\\ 
&=&   { \Delta_{\uu} \ \ \Delta_{-\vv} \over \prod_{j , k =1}^N (u_j
-v_k +i )} .  \ee
 Here and below we denote by $\Delta_{\ww} $ the Vandermonde
 determinant associated with the set of complex numbers $\ww=
 \{w_1,\dots, w_N\}$,
\be \Delta_\ww\defeq \prod_{j <k}^N (w_j -w_k ).  \ee
Then we represent the expectation value in the numerator of
\re{fermrepb} as a product of difference operators acting on the
Cauchy product  \re{Cauchydet}.  The result is
\be \la{findifex} \caS _{\uu,\vv} ={ 1\over \CK_{\uu,\vv}}\
\prod_{j=1}^N \(1 -e^{2ip_{_\uu} (v_j)}\,  e^{i \p/\p v_j} \) \( 
e^{-i \p/\p u_j} -1\)\, \CK_{\uu,\vv} , \ee
 where $e^{i \p/\p u} $ denotes the shift operator $u\to u+ i$.  We
 can commute the denominator of the Cauchy product to the left, using
 the relations
 \be \la{CommD} e^{-i \p/\p u_j} \prod _{k ,l=1}^N {1\over u_k -v_l+i}
 &=& { E^+_\vv(u_j )} \prod _{k ,l=1}^N {1\over u_k -v_l +i}\ e^{-i
 \p/\p u_j} , \no \\
  e^{i \p/\p v_j} \prod _{k ,l=1}^N {1\over u_k  -v_l+i}
 &=&  { E^-_\vv(v_j )}
  \prod _{k ,l=1}^N {1\over u_k  -v_l +i}\ 
  e^{i \p/\p v_j},
 \ee
 until it cancels its inverse.  Here and below we denote
 \be\la{defEE}
 E^\pm_\uu(v) =
     {Q_\uu(v \pm i)\over Q_\uu(v)} ,
     \quad
    E^\pm_\vv(u) =
           {Q_\vv (u \pm i)\over Q_\vv (u )}.
     \ee
 Now we can  write \re{findifex} in a factorized operator
 form,
\be \la{FactorF} \caS_{\uu,\vv} =&(-1)^N & { 1 \over \Delta_\vv }
\prod_{j =1}^N \(1 - e^{2 i p_\uu(v_j )} {E^-_{\uu}(v_j )} 
\,  e^{i \p/\p v_j}  \) \Delta_\vv 
\no \\
&& \times
 { 1 \over \Delta_{\uu} } \prod_{j =1}^N \(1 - {E^+_\vv(u_j  )} 
\, e^{-i \p/\p u_j} \) \Delta_{\uu} \ \cdot 1 .  \ee
The factorization is not complete because we must commute the
operators $e^{i \p/\p v_j }$ to the right through the functions
$E^+_\vv$ in the second factor, which depend implicitly of the
$v$-variables:
\be e^{i \p/\p v_k}\ {  E^+_\vv[u_j ]} = \(1 - {1\over
(u_j - v_k )^2 +1}\) {  E^+_\vv[u_j ]} \ e^{i \p/\p v_k}.
\ee

 \subsection{Another writing  of the operator factorization}
  
The operator representation \re{FactorF} of the inner product of an
on-shell state $\vvert \uu\, \rrangle$ and an off-shell state $\llangle
\vv\vvert$ is the main tool we are going to use to investigate the
classical limit of large $L$ and $N$.  Here we will give it a
more abstract formulation in terms of a pair of non-commuting
functional variables, which become c-functions in the classical limit.
 
For any set of points $\uu= \{u_j \}_{j =1}^N$ in the complex plane
and for any  function $f(u)$ define the functional
\be \la{defCA} \caA^\pm _\uu[f] &\defeq & { 1 \over \Delta_{\uu} }
\prod_{j =1}^N \(1 - f(u_j ) \, e^{\pm i \p/\p  u_j } \) \Delta_{\uu} 
\, ,
  \la{CAdet} \ee
 or, in terms of free fermions,
 \be \la{frefrA} 
 \caA^\pm _\uu[f] &= & {\langle N| \ \prod_{k =1}^N \[
 \bar\psi(u_k ) - f(u_k )\, \bar
 \psi(u_k \pm i) \]
 \ket \over \langle N| \prod_{k =1}^N\bar \psi(u_k )\ket}
 \no
 \\
 &=& { \det_{j k} \( u_j ^{k-1} - f(u_j ) \, (u_j \pm i)^{k-1}\)
  \over \det_{j k} \( u_j ^{k-1}\) } \, . 
  \ee
 The functionals $\caA^\pm_\uu[f]$ are completely symmetric 
 polynomials of the values of the function $f$ on the set $\uu$,
 having  total degree $N$.
 
Assuming that $\uu \cap \vv= \emptyset$, we can write the r.h.s.
of \re{FactorF} as a matrix element
\be \la{SLhatb} \caS_{\uu, \vv } &=& (-1)^N \ {( \vv |\ \caA^+_\vv
[U]\,\ \caA^-_\uu[V] \, |\uu ) \over (\vv|\uu)} \, , \ee
where the operator functions $U(v)$ and $V(u)$ satisfy the algebra
\be \la{commUV} U(v) V(u) =V(v) U(u) \, \(1 - {1\over (u-
v)^{2}+1}\)\, \ee
 and act on the vectors $(\vv|$ and $|\uu)$ as
\be
   \begin{aligned}
     U(v )\, |\uu ) &= e^{2i p_\uu(v )} \, E^-_\uu(v) 
\
     |\uu ) =  {d(v)\over a(v)}    E^+_\uu(v)  \,  |\uu ) \, , 
     \\
   (\vv| \, V(u ) &= E^+_\vv (u) \ ( \vv| .
 \end{aligned}
   \la{defUVuv} \ee

\medskip 

\noindent In this way the problem of evaluating the inner
product reduces to the problem of evaluating the functionals
\re{defCA}.

\subsection{Generalization to non-highest-weight states}

The $N$-magnon states \re{genstateket} are highest weight states with
respect to the fully ordered state, pseudo-vacuum $\ketup{L}$.  Each
operator $\bB(u_j )$ flips one spin down so that the third component
of the spin of the state with $N$ magnons built on the pseudo vacuum
$\ket$ is $\bS^z= \hf L - N$.
   
A complete system of states is obtained by acting with 
magnon creation operators on
the vacuum descendant states $(\bS^-)^{K}\ketup{L}$ with $K\le L/2$,
which correspond to ferromagnetic vacua rotated away from the
third axis.  On the other hand, the components $\bS^\pm$ of the 
total spin can be obtained by taking the infinite rapidity limit of 
the magnon creation and annihilation operators,
\be \la{Smin}
 {\bB(u) } \simeq {i\over u} \bS^-,
 \quad
  {\bC(u)  } \simeq {i\over u} \bS^+
 . 
 \ee
 The factor $1/u$ is obtained by comparing the large $u$ asymptotics
 $\braup{L}\, \bB^\dag (\bar u)\bB(u) \, \ketup{L} \simeq
 L/u^2$, which follows from \re{ddCY}, with the normalization of the
 spin operator $ \braup{L} \bS^+\bS^- \ketup{L}= L$.
Therefore an $M$-magnon state built upon a vacuum descendent can be
obtained as the limit of a $N$-magnon state \re{genstateket} with $K=
N-M$ of the rapidities sent to infinity:
\be
\begin{aligned}
\braup{L} (\bS^+)^{K' } \prod  _{j=1}^{N-K'}{\bC(v_j)   } \vert 
&=
 \llangle \vv_{_{N-K'}}\cup \infty^{K'} \vvert\, ,
\\
\prod_{l=1}^{N-K} {\bB(u_k)   } \ (\bS^-)^{K}
\ketup{L}
&= \vvert
\uu_{N-K} \cup \infty^K\rrangle,
\end{aligned}
\ee
where the limits should be taken sequentially according to 
the definition
 \be \vvert \uu_{N-1}\cup\infty\rrangle \defeq \lim _{u_N\to\infty}
 {u_N\over i} \vvert\, \uu_N \rrangle = \bS^- \vvert \uu_{N-1}
 \rrangle\, .  \ee

To evaluate the inner products of such states, we need to compute the
result of sending $u_N$ to infinity in the functionals
$\caA_\uu^\pm[f^\pm]$, assuming that the function $f^\pm(u)$ behaves at
infinity as
\be f^\pm(u) \simeq e^{ \mp i K_\pm/u}, \qquad u\to\infty.  \ee
From the definition \re{defCA} we find, taking into account that
  $\Delta_{\uu_N} \simeq u_N^{N-1}
\Delta_{\uu_{N-1}}$,
\be \la{limitABPS} \caA^\pm _{\uu_{N-1}\cup\infty}[f^\pm]&\defeq& \lim
_{u_N\to\infty} {u_N\over i} \caA^\pm _{\uu_{N}}[f^\pm] \no \\
 & = & \lim _{u_N\to\infty} {1\over \Delta_{\uu_{N-1}}} u_N^{-N+1} (1-
 e^{\mp i K/u_N} e^{\pm i \p/\p u_N}) u_N^{N-1} \
 \Delta_{\uu_{N-1}}\no \\
 &=& \pm (K_\pm - N+1)\, \CA^\pm_{\uu_{N-1}}[f^\pm].  \ee
Applying this relation the necessary number of times, one can obtain
the generalization of the operator factorization formula \re{SLhatb}
to the case of an inner product of non-highest-weight states.

\subsection{The case when the Bethe eigenstate is a vacuum descendent}

 The inner product of Bethe state with a vacuum descendent   is obtained
 from eq.  \re{SLhatb} by sending all the roots from the set $\uu$ to
 infinity.  In this limit limit the sets $\uu$ and $\vv$ are
 infinitely separated, the commutator in \re{commUV} vanishes, and
 \re{defUVuv} gives $U(v) = {d(v)\over a(v)} $ and $V(u)\sim e^{i
 N/u}$.  Applying sequentially \re{limitABPS} to all variables $\uu$,
 one obtains $ \caA^{-}_{\infty^N}[V]= (-1)^N N!$\, , and \re{SLhatb}
 gives
\be
\la{BetheVD} 
 \braup{L} \prod_{j=1}^N {\bC(v_j) } \ (\bS^-)^N \ketup{L} 
 = \caS_{\infty^N, \vv_N} = {N!} \ \caA_\vv^+[d/a]
.
\ee

To evaluate the scalar product of two vacuum descendants, we have to
send the set $\vv$ to infinity as well.  Applying sequentially
\re{limitABPS}, with $K_- = L$, to all variables from the set $\vv$,
one finds
\be \la{limBPS} \braup{L} (\bS^+)^N (\bS^-)^N
\ketup{L}=\caS_{\{\infty^N\}, \{\infty^N\}} = (N!)^2 \begin{pmatrix} L
\\
      N \end{pmatrix}.  \ee
The second factor counts the number of ways to have $N$ reversed spins
in a chain of length $L$.

\subsection{ Two-kink pseudo-vacua and restricted scalar products}
\la{sect:partiallyorderedvacua}

 In view of the applications to SYM, we are going to consider
 pseudo-vacua with two kinks at distance $K$.  In such a state the
 first $K$ spins are down and the rest $L-K$ spins are oriented up:
 \be \la{def0ml} \langle  \downarrow^K \uparrow^{L-K}\!\! | =
 \big\langle\spindown{K } \ \ \spinup{L-K}\big|    .  \ee
  This state can be obtained by acting on the left pseudo-vacuum $
  \braup{L}$ with $K$ annihilation operators \cite{Omar}
 \be \la{partialliordB} \langle
   \downarrow^K \uparrow^{L-K}\!\!|  =
\braup{L} \prod_{j =1}^{K} {\bC(z_j) }
= \llangle \zz_K \vvert \ ,
  \ee
with rapidities $\zz_{_K}= \{z_1, \dots, z_k\}$ determined by the
values of the inhomogeneity parameters associated with the first $K$
sites:
 \be \la{restrictn} z_k \equiv \th_{k}+\hf i \qquad ( k= 1,
 \dots, K).  \ee
 Hence, the inner products with the left pseudo-vacuum replaced by the
 two-kink state \re{def0ml} are evaluated by restricting $K$ of the
 rapidities\footnote{The {restricted inner product} \re{scprdN3} has
 been studied in \cite{1999NuPhB.554..647K, 2011NuPhB.852..468W,
 Omar,2012arXiv1203.5621F}.  A statistical interpretation of the
 restricted scalar product as partition functions of the six-vertex
 model is given in \cite{2011NuPhB.852..468W,2012arXiv1203.5621F}.}
    \be
    \la{scprdN3}
    \begin{aligned}
    \langle \downarrow^K \uparrow^{L-K}\!\!  | \prod_{j=1}^{N-K}
    {\bC(v_j) }\prod_{k=1}^N {\bB(u_k) }\ketup{L} =\llangle
    \vv_{_{N-K}}\cup \zz_{_K}\, \vvert \, \uu _{_{N}}
    \rrangle_{_{\thth_L}},
\end{aligned}
\ee
with 
\be \thth_{_L} = \thth_{_{L-K}} \cup \thth_{_K}\, , \qquad \zz_K
=\thth_{_K} -\hf i .  \ee

  The restricted scalar product \re{scprdN3} can be computed
  by a factorized  expression similar to \re{SLhatb},
\be \la{SLhatbz}\llangle \vv \cup \zz \, \vvert \, \uu
    \rrangle_{_{\thth } }&=& (-1)^N \ {( \vv |\ \caA^+_\vv [U]\,\
 \caA^-_\uu[V] \, |\uu ) \over (\vv|\uu)} \, , \ee
where the operator functions $U(v)$ and $V(u)$ satisfy the algebra
\re{commUV} and act on the vectors $(\vv|$ and $|\uu)$ as
   \be\la{defefrestr} \la{UVrestr}
  \begin{aligned}
    U(v )\, \vert \uu ) & \ = \k\, {1\over
    E^+_\zz(v)}{Q_\thth(v - {1\over 2} i) \over Q_\thth(v+ {1\over 2}
    i) } \ {E^+_\uu(v) } \vert \uu )\, , \\
       ( \vv\vert \, V(u ) \ &= {E^+_\zz(u )} \
       {E^+_{\vv} (u ) } \, \ ( \vv\vert \, .
   \end{aligned}
    \ee
This trivial substitution shows the power of the operator expression
\re{FactorF}.  As a comparison, when evaluating the restricted scalar
product using the original Slavnov determinant, one comes upon
spurious singularities (poles and zeroes on the top of each other),
which require multiple use of l'H\^opital's rule.

 In the limit $\uu, \vv\to\infty$, defined as in \re{limitABPS}, one
 obtains
 \be \la{BPSrestricted} \caS_{\infty^N,\,  \infty^{N-K}\cup\, \zz_K} = N!
 (N-K)!  \begin{pmatrix} L-K \\ N
\end{pmatrix}
 . \ee
The second factor in \re{BPSrestricted} counts the number of
non-equivalent ways to reverse $N$ spins among the remaining $L-K$
up-spins of the partially ordered pseudo-vacuum.

 \subsection{Gaudin-Izergin determinant and pDWPF}
 \la{ssec:GID}

In the  particular case  $K = L= N$, the restricted scalar product
\be \la{Iserg} \caZ_{\uu,\zz}\equiv  
\llangle \zz_{_N} \vvert  \uu_{_N}\rrangle_{\thth_{_N}}  
=
\bradown{N}
\prod_{j=1}^{N} {\bB(u_j ) } \ \  \ketup{N} 
   \, ,\qquad \zz_{_N} = \thth_{_N} + i/2  \, ,
 \ee
evaluates the partition function of the six-vertex model with
domain-wall boundary conditions (DWBC) on a $N\times N$ square grid
\cite{korepin-DWBC,2000nlin......8030K}.  With this specialization of
the rapidities,   the second term of the
kernel $\O(u, v)$, eq.  \re{defOhat}, vanishes at $v= z_k$ and the
Slavnov formula \re{SLhat} gives
\be \la{IzergSP} \caZ_{\uu, \zz} &=& {\det_{j k} t(u_j - z_k ) \over
\det_{j k} {1\over u_j -z_k +i} } \, ,
\quad t(u) = {1\over u} - {1\over u+i}\, .  \ee
The determinant representation \re{IzergSP} of the DWBC partition
function was obtained by Izergin \cite{Izergin-det,Izer-Cok-Kor}.  For
the first time the ratio of determinants \re{IzergSP} appeared in the
works of M. Gaudin \cite{GaudinKorDet, Gaudin-livre} as the scalar
product of two Bethe wave functions for a Bose gas with point-like
interaction on an infinite line.

The Gaudin-Izergin determinant \re{IzergSP} has an obvious writing 
in terms of free fermions, 
\be
\begin{aligned}
 \caZ_{\uu, \zz}  &= {\bra \prod_{j =1}^N \[ \psi(u_j ) - \psi(u_j
+i) \]
 \ \prod_{k =1}^N \bar\psi(z_k ) \ket \over \bra \prod_{j =1}^N
 \psi(u_j +i ) \prod_{k =1}^N\bar \psi(z_k )\ket } 
,
\end{aligned}
\la{fermrepaz}
 \ee
and is expressed in terms of the functionals \re{defCA} as
 \be \la{IzerGauddet} \caZ_{\uu, \zz} =(-1)^N
  \caA^-_\uu[E^+_\zz]       
  .    
    \ee

The limit of $ \caZ_{\uu, \zz}$ when part of the rapidities $\uu$ are
sent to infinity was recently studied by Foda and Wheeler in
\cite{FW3} and given the name partial domain-wall partition function,
or pDWPF. We compute the pDWPF by applying sequentially \re{limitABPS}
to $N-n$ of the variables, with the result
  \be
  \caZ_{\uu_n\cup \infty^{N-n},\,  \zz_N}
  = (-1)^{n}\,
    (N-n)!\,   \
  \caA^-_{\uu_n }[E^+_{\zz_N}]
\, .
 \la{repZa}  
   \ee
 An alternative proof of eq.  \re{repZa} is presented in \cite{FW3}.
The  fermionic representation of  pDWPF is
\be
\begin{aligned}
 \caZ_{\uu_n\cup \infty^{N-n},\,  \zz_N}&=
  \, (N-n)!\   \  {\langle N-n|  \prod_{j =1}^n \[ \psi(u_j ) - \psi(u_j
+i) \]
 \ \prod_{k =1}^N \bar\psi(z_k ) \ket \over\langle N-n| \prod_{j =1}^n
 \psi(u_j +i ) \prod_{k =1}^N\bar \psi(z_k )\ket } .
\end{aligned}
\la{fermrepap}
 \ee
The expectation value on the r.h.s. of \re{fermrepap} is defined for
any pair of non-negative integers $N$ and $n$, but it vanishes
identically when $n>N$, This yields a pair of identities
   \be
   \caA^\pm _{\uu_n} [E^\mp_{\zz_N}] 
   =0
   \quad \text{for}
     \ N<n
  .
   \ee

\subsection{ Properties of the functionals $ \caA^\pm _{\uu}[f]$}

\subsubsection*{Expansions }

The functionals $\caA^\pm _\uu$, defined by eq.  \re{defCA}, are
obviously completely symmetric polynomials of degree $N$ of the
variables $f(u_1), \dots , f(u_N)$.  The coefficients of the
polynomial are obtained by expanding the product in \re{defCA} as a
sum of monomials labeled by all possible partitions of the set $\uu$
into two disjoint subsets $\uu'$ and $\uu''$, with $\uu' \cup\uu'' =
\uu$,
\be \la{ExpA} \caA_{\uu}^\pm &=& \sum_{\uu'\cup \uu''=\uu }
(-1)^{|\uu'|}\ \(\prod_{u' \in\uu' } f(u' ) \) \, {1\over\Delta_{\uu}}
\prod_{u'\in\uu'}  e^{\pm i \p/\p u'} \Delta _{\uu} .  \ee
Here $|\uu'|$ stands for the number of elements of the subset $\uu'$.
The last factor is evaluated as
\be {1 \over \Delta _\uu } \(\prod_{u' \in\uu'} e^{\pm i \p/\p u'} \)
\Delta _\uu = \prod _{u' \in \uu', u''\in \uu'' }{ u' - u'' \pm i
\over u' -u'' } .  \ee
The expansion \re{ExpA}
can be used as an alternative definition of the functional
$\caA^\pm_\uu[f]$.  For $f= d/a$, this expansion was thoroughly
studied by Gromov, Sever and Vieira \cite{GSV}.  It was found in
\cite{GSV} that for constant function $f(u) = \k$, the expansion
\re{ExpA} does not depend on the positions of the rapidities $\uu$ and
the functional $\caA^\pm _\uu[f]$ is given in this case by
\be \la{triviale} \caA^\pm _{\uu}[\k ]= (1-\k)^N = \exp \( - N
\sum_{n=1}^\infty {\k^n\over n}\) .  \ee

\subsubsection*{2. The linear term in $f$ as a contour integral}

The linear term in $f$ can be evaluated as a contour integral:
\be \la{I1} \caA_\uu^\pm[f] &=& 1- \sum_{j =1}^N {f(u_j )}\prod_{k(\ne
j)}{ u_j -u_k \pm i\over u_j -u_k } + O[f^2] \no \\
& =& 1\pm \oint\limits_{A_\uu} {du\over 2\pi } \ f(u) E^\pm _\uu(u) 
 + O[f^2].  \ee
The integration contour $A_\uu$ encircles all points of the set $\uu$
and leaves outside the possible singularities of the function $f(u)$.

\subsubsection*{ Functional identities}

Using the fermionic representation \re{frefrA} and the fact that the
fermion correlator is translation invariant, we transform
 \be \la{AdetAtA} \caA^+_\uu[f]
 & =& {\langle N| \ \prod_{k =1}^N \[ \bar\psi(v_k -i ) - f(v_k )\,
 \bar \psi(v_k ) \]
\ket \over \langle N| \prod_{k =1}^N\bar \psi(v_k )\ket}.  \no \\
  & =& (-1)^N\ f(v_1)\dots f(v_N)\ {\bra \prod_{k=1}^N \[ \bar\psi(v_k
  ) - {1\over f(v_k )}\, \bar \psi(v_k -i) \]
  |N\rangle \over \bra \prod_{k=1}^N \bar\psi(v_k ) |N\rangle} \no \\
 & =& (-1)^N\ f(v_1)\dots f(v_N)\ \caA^-_\uu[1/f].  \ee
Hence, $\caA^+$ and $\caA^-$ are related by the  functional identities
 \be \la{funceqa} \caA^\pm_{\uu}[1/f]\ &=&(-1)^N {\caA^\mp _{\uu}[f]
 \over \prod_{j=1}^N f(u_j )}\, .
  \ee

 \section{Classical limit} \la{sec:classical}

\begin{figure}
         \centering
	 \begin{minipage}[t]{0.4\linewidth}
            \centering
            \includegraphics[width=3.0 cm]{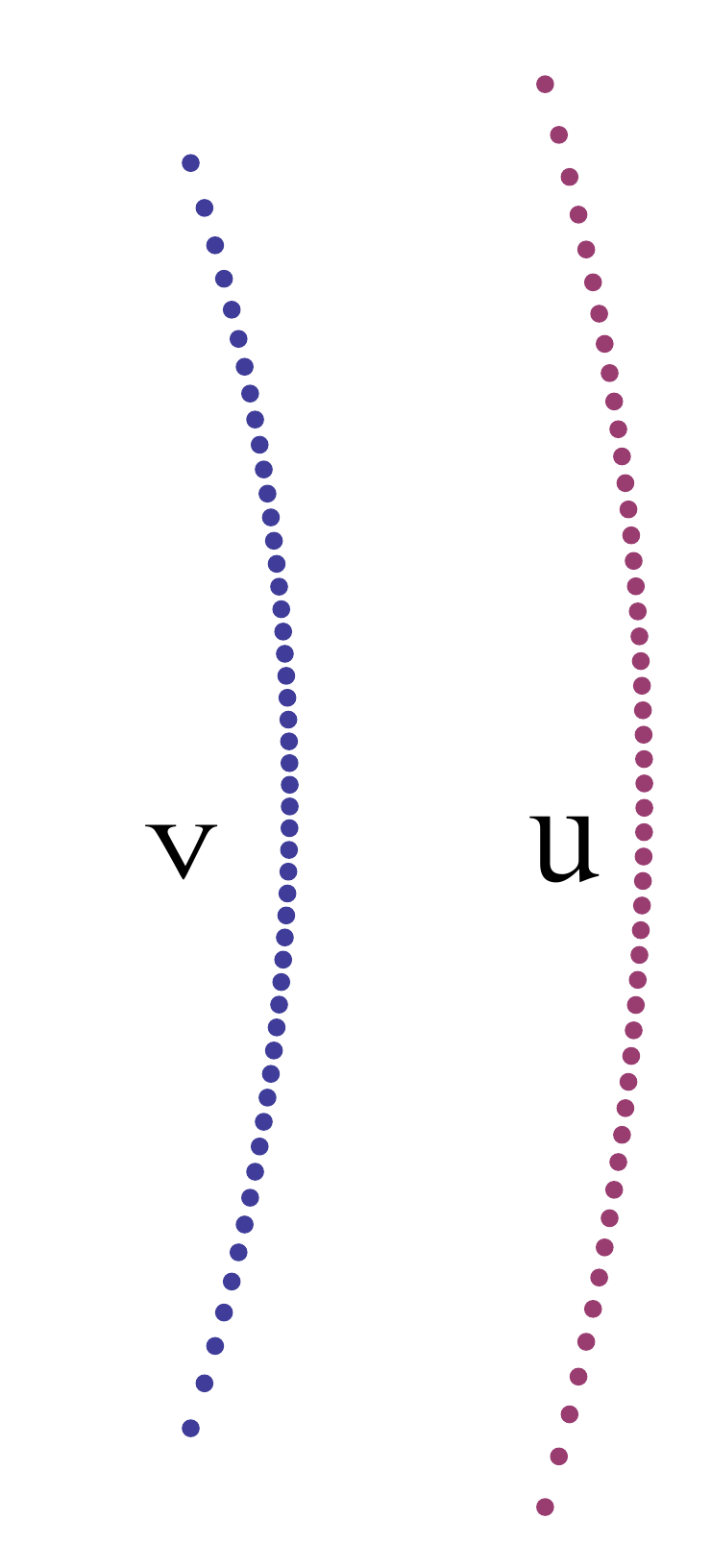}
\caption{ \small An example of the distributions $\uu$ and $\vv$ for
$N=50$, each consisting of a single macroscopic Bethe string.  }
  \label{fig:longstrings}
         \end{minipage}%
         \hspace{2cm}%
         \begin{minipage}[t]{0.4\linewidth}
            \centering
            \includegraphics[width=3.2cm]{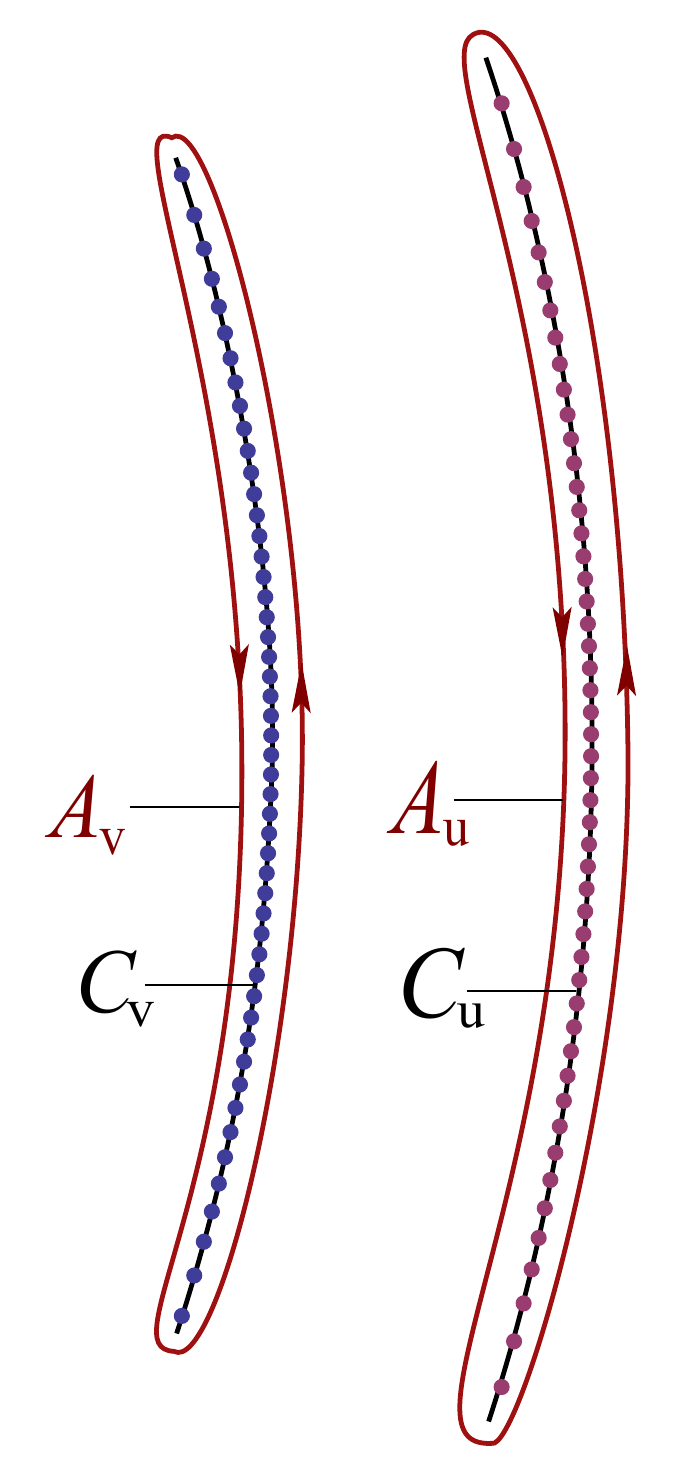}
	     \caption{ \small The cuts $C_\uu$ and $C_\vv$ and the
	     integration contours $A_\uu$ and $A_\vv$ for the one-cut
	     solution of Fig.  \ref{fig:longstrings}.}
\label{fig:RRAA}
         \end{minipage}
 \vskip 1cm
      \end{figure}

  In this section we will find the classical limit of the inner
  product \re{SLhat}.  The classical limit is achieved when $L,
  N\to\infty$ with $\a=N/L$ finite, and some additional assumptions on
  the distribution of the rapidities $\uu_N$ and $\vv_N$.  In the
  condensed matter literature the classical limit, in which each Bethe
  string has macroscopic number of particles, has been studied by
  Sutherland \cite{PhysRevLett.74.816} and by Dhar and Shastry
  \cite{PhysRevLett.85.2813}.  In this regime the roots $\uu$ condense
  along a curve $C_\uu$ in the rapidity plane, consisting of several
  connected components $C_{\uu_1}, \dots , C_{\uu_n}$, with $\uu_1\cup
  \dots \uu_n=\uu$, symmetric about the real axis, with slowly varying
  linear density $\rho_\uu(u)$ \cite{Kazakov:2004qf}.  The curve
  $C_\uu^k$ contains $N_k= \#\uu_k $ particles,
\be
  \int_{C_\uu^k} \rho(u) du= N_k\, , \qquad N_1+\dots+ N_n=N. \ee
We assume that the filling fractions $\a_k= N_k/L$ associated with the
cuts $C_k$ remain finite when $L\to \infty$.  Then the size of each
curve is $\sim L$.  We assume a similar behavior for the rapidities
$\vv$.  An example of distributions $\uu$ and $\vv$ with $N=50$ and
$n=1$ is given in Fig.  \ref{fig:longstrings}.

In the classical limit,  the  arguments $U$ and
$V$ in Eq. \re{SLhatb} become c-functions, and the inner  product
factorizes to
\be \la{SLhatbcl} \caS_{\uu, \vv } &=& (-1)^N \ \caA^+_\vv [\k \, e^{i
G_\uu- i G_\thth} ]\,\ \caA^-_\uu[ e^{i G_\vv}] \, \, ,  \ee
where 
\be G_\uu(u)= \p_u \log Q_\uu(u), \quad G_\vv(u)=\p_u\log Q_\vv(u),
\quad G_\thth(u) = \p_u \log Q_\thth (u)\ee
are the resolvents associated respectively with the sets $\uu$, $\vv$
and $\thth$.
The resolvent $G_{\uu}(u) $ is a meromorphic function of $u$ with cuts
  $C^1_\uu, \dots, C_\uu^n$ and asymptotics $N/u$ at
infinity.  The discontinuity across the cuts is proportional to the
density $\rho_\uu(u)$.  
   
 The form of the curve $C_\uu$ and the density of the distribution of
 the roots is determined by the finite-zone solution constructed in
 \cite{Kazakov:2004qf}, which we discrete shortly below.  In the
 vicinity of each cut $C_\uu^k$, the pseudo-momentum
\be
\la{defpseudomomentum}
 p(u) = G_\uu(u) - \hf G_\thth(u)+ \hf  \phi\, 
(\text{mod}\ \pi)
\ee
splits into a continuous part $ \slp(u)$, equal to the half-sum of the
values of $p(u)$ on both sides of $C_\uu$ and a discontinuous part
$\hat p(u)$, which is proportional to the density $\rho_\uu$:
    \be p_\uu(u)= \slp_\uu(u) + \hat p_\uu(u), \quad |\hat p_\uu(u)|=
    \pi \rho_\uu(u).  \ee
When the set $\uu$ satisfies the Bethe equations, the eigenvalue of
the transfer matrix
$T_\uu= 2 \cos p_\uu$ is analytic in $u$ and hence takes the same
value on both sides of the cuts $C_\uu^k$.  This yields the boundary
condition $ \sin\hat p_\uu\, \sin \sla p = 0$ on the cuts.  Along each
cut $\rho_\uu >0$, which implies $\sla p_\uu=0 \ (\text{mod}\ \pi)$,
or
 \be \la{boundcp} 
 \sla p (u) = \pi n_k, \qquad u\in C_\uu^k, \ee
where $n_k$ is the mode number associated with the cut $C_\uu^k$.  The
branch points are zeroes of the entire function $\Delta(u)\equiv T
_\uu^2(u)- 4 = - 4 \sin^2 p_\uu$.  The forbidden zones $\Delta(u)>0$
are associated with the cuts of the pseudo-momentum $p(u)$ on the
first sheet.

 The typical situation in the homogeneous limit is when $\Delta(u)$
 has a double zero at $u=\infty$, $2n$ simple zeroes $a_1,\bar a_1,
 \dots , a_n, \bar a_n$, and infinitely many negative double zeros at
 $ a_{-1}, a_{-2}, \dots a_{- k}, \dots $ where $p(a_{-k})= 2\pi
 n_{-k}$, accumulating at the point of essential singularity $u=0$.
 The cuts are along the forbidden zones between $a_k$ and $\bar a_k$
 $(k=1,\dots, n)$.  
 
 The derivative of pseudomomentum $ \p p_\uu(u)$ as well as the
 exponential $e^{2 i p(u)}$ are defined on a Riemann surface
 associated with the hyper-elliptic complex curve
\be \la{defRS} y^2 = \prod_{k=1}^n (u^{2}-a_k^{2})\, .  \ee The values
of $e^{ip(u)}$ on the first and on the second sheet are related by
\be e^{ i p^\az(u) } = e^{- i p^\za(u)}.  \ee
The derivatives of the pseudo-momentum in $N_j$,
  \be\la{abdIf} \p _{N_j} p (u) du = \o_j(u), \quad j=1,\dots, n \ee
form a basis of holomorphic abelian differentials on the complex curve
curve: 
\be\la{abdiF} \frac{1}{2\pi i}\oint_{A_\uu^k} \o_j =
\delta_{kj} , \qquad k,j=1,...,n.  \ee
The cycle  $A_\uu^k$ represents  a closed contour
on the Riemann surface  which encircles the cut $C_\uu^k$
anti-clockwise, as shown in Fig. \ref{fig:RRAA}.

By the factorization formula \re{SLhatbcl}, the computation of the
inner product in the quasiclassical limit reduces to that of the
functionals $\caA_\uu^\pm$.  This last problem was solved in the
particular case $f=d/a$ in \cite{GSV}.  Below we give an alternative,
and a little bit shorter, derivation of their result.  Let us remark
that although both derivations seem rather convincing, neither of them
is a rigorous one.

\subsection{Classical limit of the  functionals   $\caA^\pm_\uu[f]$  }

 Assume that the rapidities $\uu$ are distributed along a (possibly
 milti-component) contour $C_\uu $ in the complex $u$-plane with
 finite and sufficiently smooth density.  When the functional argument
 $f$ is small, by eq.  \re{I1},
\be\la{smallfasym} \log \caA^\pm _{\uu}[ f] &=& \pm \oint
\limits_{A_\uu} {dz\over 2\pi } \ e^{iq^\pm (z)} \ + \ O[f^2],   \ee
where the integration contour $A_\uu$ encircles 
 $C_\uu$ anticlockwise, and the function
$q^\pm(u)$ is defined by
\be \la{defq} q^\pm (u) =-i \log [ f(u)] \pm G_\uu(u) .  \ee
On the other hand, the functional relations \re{funceqa} allow us to
determine the asymptotics for large $f$, which can be written again in
the form of a contour integral.  To see that we first express
\be \la{prodf} \log\[(-1)^N \prod_{j=1}^N f(u_j )\]&=& \oint
\limits_{A_\uu } {du\over 2\pi i} \ G_\uu(u) \(\log[f(u)] +i\pi\) \no
\\
  &=& \pm \oint \limits_{A_\uu} {du\over 2\pi } \ [ \hf q_\pm ^2(u) +i
  \pi q_\pm(u)].  \ee
Substituting  \re{prodf} in  \re{funceqa}, we find the large $f$
asymptotics
\be\la{largefasym} \log \caA^\pm _\uu[ f]\simeq \oint \limits_{A_\uu}
{du\over 2\pi } \big( \pm \hf [q_\pm (u) + i\pi]^2 \mp e^{-iq^\pm (u)}
\big) \ + \ O(f^{-2}].  \ee
 We will look for a solution compatible with the behavior at $f\to 0,
 \infty$, which should be of the form
\be \la{ansatzAf} \log \caA^\pm _{\uu}[ f] = \oint _{A_\uu} {du\over
2\pi }\ F^\pm \big( e^{iq^\pm (u)}\big) , \ee
where $q(z)$ is defined by \re{defq} and the meromorphic function
$F(\o)$ has asymptotics
\be \la{asymptF} F^\pm (\o)\simeq
\begin{cases}
    \pm \o + O(\o^2) & \text{if } \ |\o|\ll 1, \\
  \mp \hf \log(-\o)^2 \mp 1/\o + O(1/\o^2) & \text{if } \ |\o|\gg 1.
\end{cases}
\ee

The function $F(\o)$ can be determined completely by comparing the
Ansatz \re{ansatzAf} with the known exact solution \re{triviale} for
$f$ constant.  Assume that the function $F^\pm(\o) $ is expanded in a
Taylor series
\be \la{TaylorF} F^\pm (\o) = \sum_{n\ge 1} F^\pm _n\ \o^{n} \ee
in some vicinity of the point $\o=0$ and compute the r.h.s. of
\re{ansatzAf} for $q^\pm (u) = - i \log \k \pm G_\uu(u)$, which
corresponds to $f(u)=\k$.  The contour integral can be evaluated by
expanding the contour to infinity, and we find
  \be \la{checkfk} \sum_{n}F_n^\pm \oint {du\over 2\pi } \ e^{\pm in
  q(u)} = \sum_{n}F_n^\pm \oint {du\over 2\pi } \ \k^n\, ( 1 \pm i
  {N\over u} + \dots)^n = \mp \sum_n F_n^\pm \, nN \k\!^n .  \ee
Comparing \re{checkfk} and \re{triviale} we find that $F^\pm _n= \pm
1/n^2$ and that the Taylor expansion \re{TaylorF} is that of the
dilogarithm,
\be F^\pm (\o) = \pm \sum_{n=1}^\infty \ {\o^n\over n^2} = \pm\
\Li(\o) .  \ee
The asymptotic behavior \re{asymptF} is satisfied thanks to the
functional equation for the dilogarithm,
\be \la{Lifunceq} \Li({1\over \o} )= - \Li(\o) - {\pi^2\over 6} -
{1\over 2} \log^2(-\o).  \ee
Moreover, the property \re{Lifunceq} of the dilogarithm leads to a
pair of functional equations for $\caA_\uu^\pm[f]$, which are the
scaling limit of \re{funceqa}.

If the resolvent $G_\uu$ has several cuts on $C_\uu^1,\dots, C_\uu^n$,
then the integration contour in \re{ansatzAf} splits into $n$ disjoint
contours $A_\uu^1,\dots, A_\uu^n$, and the functional $\caA^\pm
_{\uu}[f]$ is given in the classical limit by
   \be \la{scalingA} 
   \log\caA^\pm_\uu [ f] & \simeq&    \pm \oint
   \limits_{ A_\uu} \frac{du}{2\pi } \ \text{Li}_2\big(f(u)\, e^{\pm
   iG_\uu(u)}\big) \, , \quad A_\uu = \cup_{k=1}^nA^k_\uu.  \ee
 The $k$-th term grows as $\a_k L$, where $\a_k= N_k/L$ is the filling
 fraction associated with the cut $C_\uu^k$.  Let us emphasize that in
 the derivation of \re{scalingA} we did not assume that the set $\uu$
 is on-shell.

 Finally, let us note that the functional identity \re{funceqa}, or
 equivalently, the property of the dilogarithm \re{Lifunceq}, leads to
 a second integral representation,
 \be \la{scalingA1}
 \begin{aligned}
\log \caA^\pm_\uu [ f] 
 &\simeq \mp \oint \limits_{ A_\uu} \frac{du}{2\pi } \
 \text{Li}_2\big(f^{-1}(u)\, e^{\mp iG_\uu(u)}\big) + \oint_{A_\uu}
 {du\over 2\pi} G_\uu(u) \log f(u)\, + i\pi N. \\
\end{aligned}
 \ee

 \subsection{ Classical limit of the Slavnov inner product
 $\caS_{\uu,\vv}$ }
 
 Substituting \re{scalingA1} in \re{SLhatbcl}, we write the logarithm
 of the scalar product as a contour integral
\be 
\la{intexplnreg} \log \caS _{\uu, \vv}= i \pi N+ \oint
\limits_{A_\vv} \frac{du}{2\pi } \ \text{Li}_2\left( \k\, e^{ i
G_{_\uu}(u) + i G_\vv(u) - iG_\thth(u) }\right) - \oint
\limits_{A_\uu} \frac{du}{2\pi } \ \text{Li}_2\left( e^{ - i
G_{_\uu}(u) + i G_\vv(u)}\right) .
\ee
The r.h.s. of \re{intexplnreg} can be reformulated in terms of a
contour integral around the ensemble of the cuts of the function
  \be \la{defqphvv} q(u) &\defeq & G_\uu(u) + G_\vv (u)- G_\th (u) 
  + \phi
  \ee
on the physical sheet of its Riemann surface, with integrand depending
only on $q(u)$.  This follows from the fact that the resolvent $G_\uu$
satisfies on its cuts $C^k_\uu$ the boundary condition \re{boundcp},
\be 2 \sla G_\uu(u ) - G_\thth(u) +\phi = 2\pi n_k \ \ \text{for}
\ \ u\in C_\uu^{k}.  \ee
(Here $\sla G_\uu$ is the half-sum of the values of the resolvent on
both sides of $C^k_\uu$ and $n_k\in \IZ$ is the corresponding mode
number.)  Hence, if $q^{(1)} $ is the value of the function $q(z)$ on
the physical sheet, defined by \re{defqphvv}, then the value of $q(u)$
on the second sheet under the cut $C_\uu^k$ is given by $q^{(2)} = -
G_\uu+ G_\vv $.  We conclude that the two integrals in
\re{intexplnreg} have the same integrand $\Li(e^{i q})$, but the
contours $A_\uu$ are placed on the second sheet of the Riemann surface
of the function $q(u)$.  After pulling all connected components
$A^k_\uu\subset A_\uu$ up to the first sheet across the cuts
$C_\uu^k$, eq.  \re{intexplnreg} takes the form
 \be \la{intrepf} \log \caS_{\uu,\vv }&=& i \pi N+\oint \limits_{A
 _\uu \cup A_\vv } \frac{du}{2\pi } \ \text{Li}_2(e^{i\, q(u)}).  \ee
(The minus sign is compensated by the change of the orientation of the
contours $A_\uu^k$ after they are moved to the first sheet.)

\begin{figure}
	 \centering
	 \begin{minipage}[t]{0.4\linewidth}
	    \centering
	    \includegraphics[width=6cm]{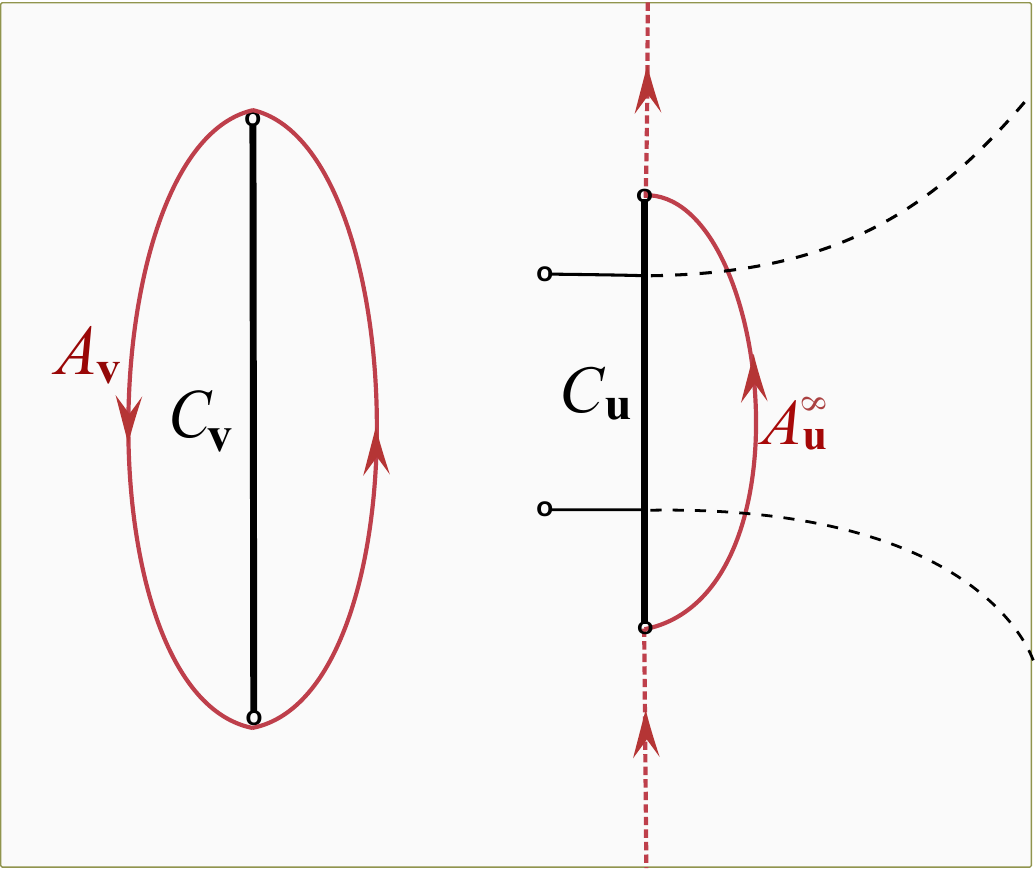}
 \caption{ \small The contour $A_\vv $ and the deformed contour $
 A_\uu^\infty$ for the integral in \re{intrepf}.  The contour $
 A_\uu^\infty$ starts at $z= - i\infty$ on the second sheet, passes on
 the first sheet at the lower branch point, continues to the upper
 branch point, where it returns to the second sheet and continues to
 $z=+i\infty$.  }
  \label{fig:Contour1}
         \end{minipage}%
         \hspace{2cm}%
	 \begin{minipage}[t]{0.4\linewidth}
	    \centering
	    \includegraphics[width=3.3cm]{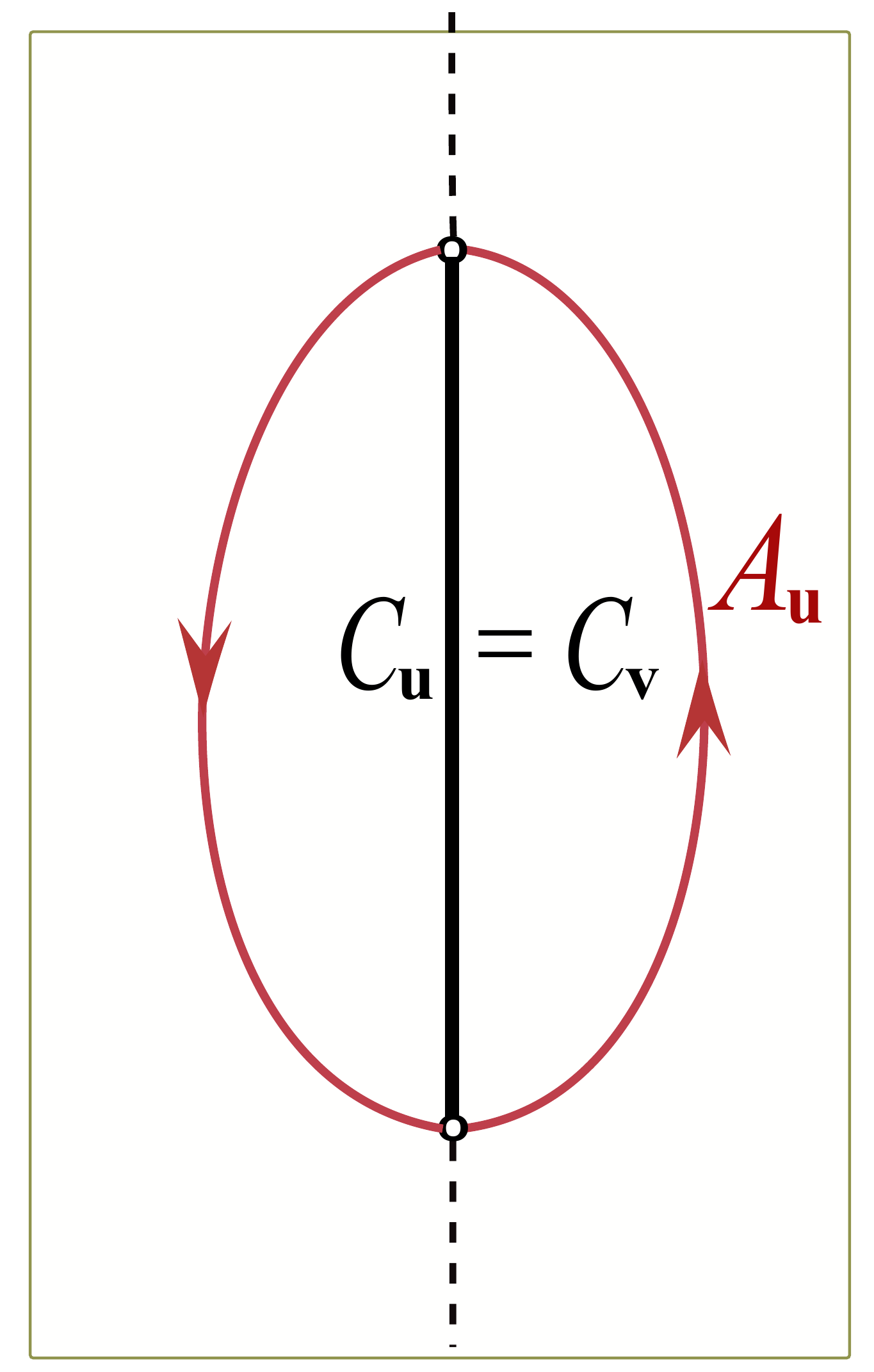}
	     \caption{\small The contour $A_\uu$ for the integral in
	     \re{classnorm}, obtained as the limit $\vv\to\uu$ of the
	     integral \re{intrepf}.  When $C_\vv\to C_\uu$, the two
	     logarithmic branch points on the first sheet join the two
	     simple branch points at the ends of the cut $C_\uu$.  }
\label{fig:ContourUU2}
	 \end{minipage}
 \vskip 1cm
      \end{figure}

The integral along $A^k_\uu$ is however ambiguous, because the
integrand has two logarithmic cuts which start at $u=\infty$ on the
second sheet, cross the cut $C_\uu^k$ and end at two branch points on
the first sheet.  The ambiguity is resolved by deforming the contour
$A_\uu $ to a contour $A^\infty_\uu$ which encircles also the point
$z=\infty$ on the second sheet.\footnote{The author is indebted to
Nikolay Gromov for performing a numerical test and for suggesting how
to place the integration contours.} In the case of a one-cut solution,
the contour $A_\uu ^\infty$ is depicted in Fig.  \ref{fig:Contour1}.

In the limit $\uu\to\infty$, eq.  \re{intrepf} must reproduce the
result of \cite{GSV} about the scalar product of two general Bethe
states and a vacuum descendent (eqs.  (3.28)-(3.29) of \cite{GSV}).
This is indeed the case.  In the limit $\uu\to\infty$, the integration
goes only along the contour $C_\vv$ and the function $q$ in the
integrand is given in the homogeneous limit by $q= G_\vv- {L\over u}
+\phi $.

\subsection{Classical limit of the Gaudin norm  }

An expression for the square of the Gaudin norm in the classical limit
can be formally obtained from \re{intrepf} by taking $G_\vv=G_\uu$.
Here we assumed that the two sets of rapidities are invariant under
complex conjugation, $\bar \uu=\uu$ and $\bar\vv=\vv$.  When $\vv\to
\uu$, the integration contour $A_\uu$ in \re{intrepf} can be closed
around $C_\uu$ and $C_\vv$ as in Fig.  \ref{fig:ContourUU2}, and in
the integrand one can replace $q(z)\to 2p(z)$, where $p(u)$ is the
quasi-momentum \re{defpseudomomentum}.  We find for the square of the
Gaudin norm $\llangle \uu\vvert\uu\rrangle ^2 = \caS_{\uu,\uu}$, with
 \be \la{classnorm} \log \caS_{\uu, \uu } &=& \oint \limits_{ A_{\uu} }
 \frac{du}{2\pi } \ \Li \( e^{ 2i p(u)}\).  \ee
  The term $i \pi N$ is compensated by another such term which appears
  because of the hermitian conjugation $\CB^\dag = - \CC$.
  Furthermore, when $C_\vv^k=C_\uu^k$, the two cuts on the second
  sheet end at the two branch points where $1- e^{2i p_\uu}= 2 i
  e^{ip_\uu}\sin p_\uu =0$ and $\Li(1-e^{2i p_\uu})$ has a logarithmic
  singularity $\Li \sim \sin (p_\uu) \log \sin (p_\uu)$.  Therefore
  there is no obstructions for placing the integration contour
  $A_\uu$.

An expression of the Gaudin norm as a linear integral was derived in
\cite{GSV}.  One can check, using the fact that $p_\uu(z) = \pm i\pi
\rho_\uu(z)$ on the two edges of the cut, that the contour integral
\re{classnorm} can be transformed into (twice) the linear integral in
eq.  (2.15) of \cite{GSV}.

\subsection{Classical limit of the restricted inner  product }

From Eqs. \re{SLhatbz}-\re{UVrestr}, one readily evaluates the classical
 limit of the restricted scalar product
\re{scprdN3},  
  \be \la{SLhatbClR} \caS_{\uu, \vv\cup \zz } &=& (-1)^N \
  \caA^+_{\vv\cup\zz} [ \k\, e^{i G_\uu - iG_\thth -iG_\zz} ]\,\
  \caA^-_\uu[ e^{i G_{\vv } + iG_\zz}] , \ee
where $\thth\cup\zz$ are the inhomogeneity parameters, 
and then use \re{scalingA}.  We find  
  \be \la{intexplnregdd} \log \caS _{\uu, \vv\cup \zz}\!\!  &=& \!\!
  i\pi N+\!  \oint \limits_{A_{\vv}} \frac{du}{2\pi } \
  \text{Li}_2\left( \k \, e^{ i G_{_\uu}(u) + i G_{ \vv}(u) - iG_\thth
  (u) }\right)\!  - \!  \oint \limits_{A_\uu} \frac{du}{2\pi } \
  \text{Li}_2\left( e^{ - i G_{_\uu}(u) + i G_\vv(u) + i G_\zz
  (u)}\right) \no \\
&=& i\pi N+ \oint \limits_{A_\uu\cup A_{ \vv} } 
 \frac{du}{2\pi } \
\text{Li}_2 \( e^{i
G_\uu(u) +i G_{ \vv} (u) - i G_\thth(u) } \).  \ee
In the second line we used the classical Bethe equation
for $\phi=0$
\be
2 \sla G_\uu(u) 
-  G_\thth(u) - G_\zz(u) = 0 \, (\text{mod}\, 2\pi), \quad u\in C_\uu.
\ee
We see that the restricted scalar product $\caS_{\uu,\vv\cup\zz}$ is
given by the same contour integral \re{intrepf}, where the integrand
depends only on $\uu, \vv$ and $\thth$.  The dependence on $\zz$ is
through the boundary condition on the cuts of the resolvent $G_\uu$.


\subsection{The derivatives in $N_k$}

Knowing the hyper-elliptic Riemann surface \re{defRS}, it is possible
to compute the logarithmic derivatives of the scalar product with
respect to the filling numbers $N_k$.  For that we use the fact that,
according to \re{abdIf}, the derivatives of the pseudo-momentum form a
basis of abelian differentials for the Riemann surface.

 Take for example the Gaudin norm.  From \re{classnorm} we find for
 its logarithmic derivative
 \be \la{classnormmm}
 \begin{aligned}
  {\p \log \caS_{\uu, \uu } \over \p N_k} &=- 2i \oint \limits_{
  C_{\uu} } \frac{dz}{2\pi } {\p p_\uu\over \p N_k} \ \log \( 1- e^{
  2i p_\uu(z)}\)
    \\
 &= - {i\over \pi} \oint \limits_{ C_{\uu} } \o_k(z) \ \log \( 2 \sin
 p_\uu(z)\) .
   \end{aligned}
     \ee
  The differentials $\o_k(u) = \p_{N_k} p_\uu(u)du $ are analytic on
  the hyper-elliptic curve \re{defRS} and are obtained as a solution
  of the relations \re{abdiF}.

 \section{\label{sec:3pf}3-point functions of trace operators in
 $\CN=4$ SYM}

 \subsection{The three-point function}

In a $su(2)$ sector of the SYM theory, the operators are made of two
complex scalars.  We consider,  as in \cite{EGSV, Escobedo:2011xw,
GSV}, the correlation function of three single-trace operators of the
type
\be \la{constitution}
\begin{aligned}
\CO_1&\sim \Tr[ Z^{L_1 - N_1} X^{N_1} + \dots],
\\
 \CO_2 & \sim \Tr[\bar Z^{L_2- N_2}\bar X^{N_2}+ \dots ],
\\
 \CO_3 &\sim \Tr[Z^{L_3- N_3}\bar X^{N_3}+ \dots],
\end{aligned}
 \ee
 where the omitted terms are weighted products of the same
 constituents taken in different order.  The weights are chosen so
 that the operator $\CO_i$ is an eigenstates of the dilatation
 operator with dimensions $\Delta_i$.  The two-point and the
 three-point functions are determined, up to multiplicative factors,
 by the conformal invariance of the theory,
\be \la{twopf}
 \< \CO_i(x_i) \CO_j(x_j)\> = L _i\ \d_{i, j}\ { \CN_i
\over |x_1-x_2|^{2\Delta_i }}, \ee
\be \la{3ptfCFT} 
\< \CO_1(x_1)\CO_2(x_2)\CO_3(x_3)\>=  {L_1L_2L_3\ 
\sqrt{  \CN_1  \CN_2  \CN_3 } \ \ \ 
C_{123}(\l) \over |x_{12}|^{\Delta_1 +\Delta_2-\Delta_3}
|x_{23}|^{\Delta_2 +\Delta_3-\Delta_1} |x_{31}|^{\Delta_3
+\Delta_1-\Delta_2} },  \ee
where $\CN_i$ are arbitrary normalization factors.\footnote{Our
normalization is slightly different that the normalization used in
\cite{EGSV}, namely $\CN_i^{\text{here}} = \CN_i^{\text{there}}/{L_i}$
and $C_{123}^{\text{here}} =C_{123}^{\text{there}} /\sqrt{L_1L_2L_3}$.
} The factor $L_i$ count for the cyclic rotations of the trace
operator $\CO_i$.  The structure constant $C_{123}(\l) $ has
perturbative expansion
\be 
N_c\ C_{123}(\l)  = C^{(0)}_{123} + \l\ C^{(1)}_{123} +\dots\, ,
\ee
where $N_c$ is the number of colors and $\l$ is the 't Hooft coupling.

  \subsection{The structure constant in terms of scalar products of
  Bethe states}

\begin{figure}
 \vskip - 1cm
         \centering
                  \begin{minipage}[t]{0.8\linewidth}
            \centering
          \includegraphics[width=11cm]{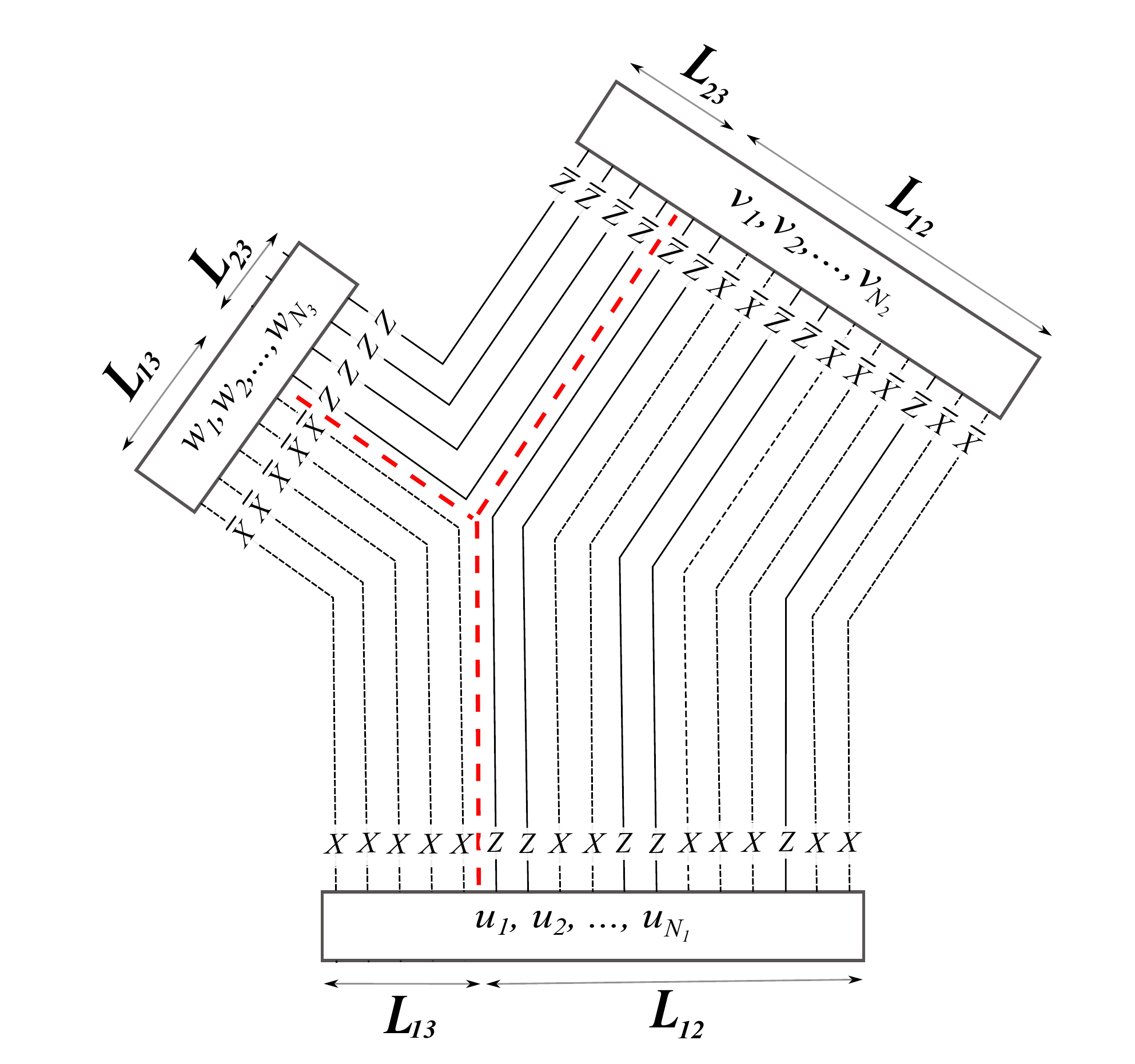}
\caption{ \small An example of a planar set of contractions
contributing to the three-point function of the operators $\CO_1$,
$\CO_2$ and $\CO_3$.  The $Z-\bar Z$ and the $X-\bar X$ propagators
are represented respectively by continuous and dashed lines.  }
  \label{fig:3pt}
         \end{minipage}
      \end{figure}

At tree level, the structure coefficient is a sum over all possible
ways to perform the Wick contractions between the fundamental fields
and their conjugates.  A non-zero result is obtained only if
 \be
 N_1 = N_2 + N_3
 \ee
 and the number of contractions $L_{ij}=\hf(L_i-L_j-L_k)$ between the
 operators $\CO_i$ and $\CO_j$ are
 \be
 L_{12}= L_1-N_3, \ \ \ L_{13}= N_3, \ \ \ L_{23}= L_3-N_3.
 \ee
The product of all free propagators in the contractions between
$\CO_i$ and $\CO_j$ reproduce the factor
$|x_{ij}|^{-\Delta_i-\Delta_j+\Delta_k}$ in \re{3ptfCFT}, with
tree-level conformal dimensions $\Delta_i = \Delta_i^{(0)}= L_i$.  By
planarity, all $Z$ fields in $\CO_3$ must contract with $\bar Z$
fields in $\CO_2$ and all $\bar X$ fields in $\CO_3$ must contract
with $X$ fields in $\CO_1$, as shown in Fig.  \ref{fig:3pt}.  Hence
there is a single term in the sum in Eq.  \re{constitution},
$\Tr(Z^{L_{23}}\bar X^{N_{3}} )$, for which the planar contractions
with $\CO_1$ and $\CO_3$ do not vanish.

In order to compute the tree-level structure constant $C_{123}^{(0)}$
by the method of \cite{EGSV}, one should know the wave functions at
one loop.  At one loop level, the operator $\CO_i$ is represented by a
$N_i$-magnon Bethe eigenstate\footnote{For simplicity consider
highest-weight states.  The generalization to arbitrary states is
outlined in section \ref{sect:partiallyorderedvacua}.} with energy
$\Delta_i$ of the periodic XXX$_{1/2}$ spin chain of length $L_i$
($i=1,2,3$).  Let us choose the pseudovacua in the three chains as \be
\Tr (Z^{L_1}) \ \to \vvert \uparrow ^{L_1}\rrangle\, , \ \Tr( \bar
Z^{L_2}) \ \to \vvert \uparrow ^{L_2}\rrangle\, , \ \ \Tr (Z^{L_3}) \
\to \vvert \uparrow ^{L_3}\rrangle\,.  \ee
Then the three operators are determined by three Bethe eigenstates
\be \la{mapOpBethe} \CO_1 \to \vvert \uu \rrangle_{_{L_1}}, \quad
\CO_2\to \vvert \vv\rrangle_{_{L_2}}, \quad \CO_3\to \vvert
\ww\rrangle_{_{L_3}}, \ee
with $\uu=\{ u_1,\dots, u_{N_1}\}$, $\vv=\{v_1,\dots, v_{N_2}\}$,
$\ww=\{w_1, \dots, w_{N_3}\}$.  
A  natural   normalization of the
two-point function is given by choosing
\be\la{normal2pf} \CN_1= \llangle \uu\vvert\uu\rrangle, \quad \CN_2=
\llangle \vv\vvert\vv\rrangle , \quad \CN_3= \llangle
\ww\vvert\ww\rrangle .  \ee
The tailoring (= cutting+flipping+sewing) 
prescription of \cite{EGSV} gives 
\be \la{C123} C_{123}^{(0)} = {
 \llangle \vv \, \vvert
\uu\rrangle_{L_{12}} 
\, \, \llangle \downarrow^{N_3} \, \vvert\, \ww
\rrangle_{L_{13}} 
\over \sqrt{\llangle
\uu\vvert\uu\rrangle_{L_1} \llangle
\vv\vvert\vv\rrangle_{L_2} \llangle
\ww\vvert\ww\rrangle_{L_3}}
}   .
  \ee

 Since the inner products in the numerator are evaluated for
 subchains, the Bethe vectors are no more on shell.  However, as
 proposed in Ref.  \cite{Omar}, one can replace $\llangle
 \vv\vvert\uu\rrangle_{L_{12}}$ by an inner product of the type
 $\llangle \dots \vvert\uu\rrangle_{L_{1}}$, which can be evaluated by
 the Slavnov formula.  This can be done after deforming the problem by
 introducing impurities with rapidities $ {\thth ^{(i)}} =\{\th_l
 ^{(i)}\}_{l=1}^{L_i}$ at the sites of the $i$-th spin chain
 $(i=1,2,3)$.      We denote the rapidities
 associated with the contractions between the operators $\CO_i$ and
 $\CO_j$ by $\thth^{(ij)}$, so that $\thth^{(1)}= \thth^{(12)}
 \cup\thth^{(13)}$, etc.
Along the lines of  \cite{Omar} we obtain for the 
structure constant, up to a phase factor, 
\be \la{C123} 
C_{123}^{(0)} = { \llangle \vv \cup \zz \, \vvert
\uu\rrangle_{\thth^{(1)}} \, \, \llangle\zz\, \vvert\, \ww
\rrangle_{\thth^{(3)}} \over \sqrt{\llangle
\uu\vvert\uu\rrangle_{\thth^{(1)}} \llangle
\vv\vvert\vv\rrangle_{\thth^{(2)}} \llangle
\ww\vvert\ww\rrangle_{\thth^{(3)}}}} 
\qquad \text{ with} 
 \ \ \  \zz = \thth^{(13)}+ i/2.
 \ee
 In this way we expressed the structure constant in terms of the 
 quantities evaluated in Sections \ref{sect:partiallyorderedvacua}
 and \ref{ssec:GID}.

 \subsection{The BPS limit}

The structure coefficient should be normalized so that in the limit
when all rapidities go to infinity it tends to the structure
coefficient for three BPS fields \cite{1998hep.th....6074L}, \be \lim
_{\uu,\vv,\ww\to\infty } C^{(0)}_{123} = C^{\text{BPS}}_{123}.  \ee
 From \re{BPSrestricted} we obtain the expected result
 \be
 \begin{aligned}
  C^{\text{BPS}}_{123} &= { { N_1!(N_1-N_3) !  \begin{pmatrix} L_{1} -
  N_3 \\
      N_2
\end{pmatrix}
\times N_3!  \begin{pmatrix} L_{3} - N_3 \\
      N_3
\end{pmatrix}
 \over \sqrt{ (N_1!)^2 \begin{pmatrix} L_1 \\
      N_1
\end{pmatrix}
\times (N_2!)^2
\begin{pmatrix}
      L_2 \\
      N_2
\end{pmatrix}
\times (N_3!)^2 \begin{pmatrix} L_3 \\
      N_3
\end{pmatrix}
}} } = { \begin{pmatrix} L_{12} \\
      N_2
\end{pmatrix}
\ \begin{pmatrix} L_{23} \\
      N_3
\end{pmatrix}
\over \sqrt{
\begin{pmatrix}
      L_1 \\
      N_1
\end{pmatrix}
\begin{pmatrix}
      L_2 \\
      N_2
\end{pmatrix}
\begin{pmatrix}
      L_3 \\
      N_3
\end{pmatrix}
}}.
\end{aligned}
\ee

\subsection{The   limit of three classical operators}
 
 Here we take the limit when the three fields become classical,
 \be L_i\to\infty \quad \text{with} \ \ \ \a_i = {N_i\over L_i}\ \ \
 \text{fixed} \qquad (i=1,2,3).  \ee
using the results of Section \ref{sec:classical}.  The two factors in
the numerator in \re{C123} are evaluated using 
\re{intexplnregdd},
 \be\la{factorUV1}
\log   \llangle \vv \cup \zz \, \vvert
\uu\rrangle_{\thth^{(1)}}
 &=& i\pi N_1+ \oint \limits_{A_\uu\cup A_{
 \vv} } \frac{du}{2\pi } \ \text{Li}_2 \( \exp i [G_\uu +  G_{ \vv} -  
 G_{\thth^{(12)}} ] \) \\
 \no
&=& i\pi N_1+ \oint \limits_{A_\uu\cup A_{ \vv} } \frac{du}{2\pi } \
\text{Li}_2 \( \exp i [p_\uu +  p_{ \vv} + \hf   G_{\thth^{(3)}}] \)\, ,
 \\
 \log  \llangle\zz\, \vvert\, \ww
\rrangle_{\thth^{(3)}} &=& i\pi N_3 +  \oint \limits_{ A_\ww } 
\frac{du}{2\pi }\ 
 \text{Li}_2\!\left( \exp  i[ G_\ww - G_{\thth^{(23)}}] \right)
\\
\no
 \la{factorW}
&=& 
 i\pi N_3 + \oint \limits_{ A_\ww } 
\frac{du}{2\pi }\ 
 \text{Li}_2\!\left( \exp  i[ p_\ww  +\hf    G_{\thth^{(2)}} 
 -\hf    G_{\thth^{(1)}}  ] \right)
 ,
\ee
where we introduced the three pseudo-momenta (for $\phi=0$)
\be \la{pseudo123} p_\uu = G_\uu - \hf G_{\thth ^{(1)}} ,\ \ \ p_\vv=
G_\vv - \hf G_{\thth ^{(2)}} ,\ \ \ p_\ww = G_\ww - \hf G_{\thth
^{(3)}}. \ee
The norms in the denominator are evaluated using \re{classnorm}.
Collecting all terms we find
\be \log \la{clasC123} { C^{(0)}_{123} } &\simeq& \oint \limits_{
A_\uu\cup A_\vv } \frac{du}{2\pi } \ \text{Li}_2\big( e^{ i[p_\uu +
p_\vv + {1\over 2}G_{\thth^{(3)} } ]}\big) + \oint \limits_{A_\ww}
\frac{du}{2\pi }\, \text{Li}_2\!\left( e^{ i[p_\ww + {1\over 2}
G_{^{\thth^{(2)}}} -{1\over 2} G_{^{\thth^{(1)}} }]} \) \no \\
& -&\hf \int\limits_{A_\uu}{du\over 2\pi} \, \text{Li}_2\big( e^{2 i p_\uu}
\big)- \hf\int\limits_{A_\vv}{du\over 2\pi} \, \text{Li}_2\big( e^{2 i p_\vv}
\big) - \hf\int\limits_{A_\ww}{dz\over 2\pi} \, \text{Li}_2\big( e^{2 i
p_\ww} \big) .  \ee
The tree-level structure constant is obtained by setting all 
inhomogeneity parameters to zero:
\be \hskip -0.5cm \la{clasC123bis}
\begin{aligned}
   \log C^{(0)}_{123} &\simeq \oint \limits_{ A _\uu\cup A_\vv }
   \frac{du}{2\pi } \ \text{Li}_2\big(e^{ i p_{\uu}(u) + i p_\vv(u)+i
   { L_3/2u}} \big) + \oint \limits_{A_\ww} \frac{du}{2\pi }\
   \text{Li}_2\left( e^{   i p_\ww(u)+i { (L_2- L_1)/2 u} } \right) \\
& -\hf \int_{A_\uu}{du\over 2\pi} \ \text{Li}_2\big( e^{2 i p_\uu(u)}
\big) - \hf\int_{A_\vv}{du\over 2\pi} \ \text{Li}_2\big( e^{2 i
p_\vv(u)} \big) - \hf\int_{A_\ww}{du\over 2\pi}\ \text{Li}_2\big( e^{2
i p_\ww(u)} \big).
  \end{aligned}
  \ee
To make connection with the result of \cite{GSV}, one should send
all $u$'s to infinity, which is the same as taking $G_\uu=0$ and
neglecting the integration along $A_\uu $.

\section{Conclusions and speculations}

The principal result of this work is the operator factorization
formula \re{SLhatb} for the scalar product and its clasicial limit
\re{smallfasym}-\re{SLhatbcl}.  Using this result, we were able to
write a compact expression for the correlation function of three
non-BPS operators in maximally supersymmetric Yang-Mills theory in the
classical limit when the traces become large.  Our starting point was
Foda's determinant expression for the three-point structure constant
\cite{Omar}.  The determinant formula of \cite{Omar} was derived
supposing that the three operators are deformed by a set of
inhomogeneity parameters, whose values can be chosen at will.  We
computed the bosonized determinant expression for inhomogeneous
problem and took, as in \cite{Omar}, the homogeneous limit at the very
end in order to avoid spurious singularities.

Another reason to treat the inhomogeneous problem is that this allows
to extend, as argued in \cite{2012arXiv1202.4103G, Didina-Dunkl}, the
tree-level result to higher orders in $\l$.  Gromov and Vieira
\cite{2012arXiv1202.4103G} showed that knowing the tree level solution
for $C^{(0)}_{123}$ in presence of impurities, one can obtain the
one-loop and the two-loop corrections by applying a special
differential operator acting on the inhomogeneity parameters
$\thth^{(1)}$, $\thth^{(2)}$ and $\thth^{(3)}$.  Serban
\cite{Didina-Dunkl} proposed that this statement can be extended to
all loops in the BDS model \cite{Beisert:2004hm}, i.e. when the
dressing phase is not taken into account, and in the limit of large
lengths $L_1,L_2,L_3$.  The analysis of \cite{Didina-Dunkl} leads to
the prescription that the higher loop corrections can be taken into
account only by modifying the pseudo-momenta.  For example, the
three-loop result for the structure constant is obtained by changing
the pseudo-momenta $p_\uu, p_\vv$ and $p_\ww $ according to the
three-loop Bethe ansatz equations \cite{Beisert:2004hm}.

According to \cite{Didina-Dunkl},  for finite value of the 't Hooft
coupling $\l$, the structure constant for three classical operators in
the BDS model is given by \re{clasC123} with a particular distribution
of the $L_i$ inhomogeneity parameters\footnote{ This is the
distribution of $L_i$ equal charges confined to the segment $[-2g,
2g]$ in absence of external potential.} in the interval $[-2g, 2g]$,
where $\l = 16 \pi^2g^2$.  For this distribution the resolvents for
the inhomogeneities associated with the three chains are given by
\be \la{thetas} G_{\thth^{(i)}}(u) = {L_i\over \sqrt{u^2-4g^2}} = {
L_i \over x} {dx\over du} \qquad (i=1,2,3), \ \ee
where $x $ is the  'Zhukovsky variable'  defined by
\be
\la{Zhuk}
u= x + g^2/x.
\ee
This has the same effect as changing the vacuum eigenvalues $a(u)$ and
$d(u)$ of the transfer matrix, eq.  \re{defad}, to
 \be \la{newad} a(u) = [x(u+\hfi)]^L,\ \ d(u) = [x(u-\hfi)]^L. \ee
In the strong coupling regime it is convenient to perform the change
the variable \re{Zhuk} in the contour integrals in \re{clasC123}.  We
should mention here that the replacement \re{newad} as a possible way
to take into account (some of) the loop corrections was previously
discussed in \cite{Escobedo:2011xw}.

Assuming that this conjecture is correct, the expression \re{clasC123}
with the choice \re{thetas} for the inhomoheneities will give the
all-loops result for the structure constant for the BDS model in which
the dressing phase is neglected.  On the other hand, the effect of the
dressing phase \cite{Beisert:2006ez}, in the limit of large
$L_1,L_2,L_3$, is that the pseudo-momentum is modified as
\be
\la{quasiBES}
p(u) \to p^{\text{BES}}(u) =
p  (u) - i \log \s^{\text{BES}}(u).
\ee
 The fact that \re{clasC123} depends only on the three quasi-momenta
 $p_\uu, p_\vv, p_\ww$ and the resolvents $G_{\thth^{(i)}} = L_i
 \p_u\log x$ invites one to consider the possibility that the result
 at finite 't Hooft coupling $\l$ is given in the classical limit
 again by \re{clasC123}, with the three pseudo-momenta modified
 according to \re{quasiBES}.

It is natural to expect that a systematic approach to the correlation
functions of heavy operators in the full field strength multiplet of
$\CN = 4$ SYM should be some extension of the algebraic curve method
used in the spectral problem in \cite{Kazakov:2004qf, Kazakov:2004nh,
Beisert:2005bm}.\footnote{Very recently, Janik and Laskos-Grabowski
\cite{2012arXiv1203.4246J} showed that the algebraic curve formalism
can be used to compute Wilson loops and the correlators between a
Wilson loop and a local operator.} In the case of three long-trace
operators, the structure constant $C_{123}$ is expected to be
described by the ensemble of three algebraic curves, associated with
the pseudo-momenta $p_\uu, p_\vv,p_\ww$.  In order to build a general
algebraic curve formalism, one should learn how to compute more
general inner products, at least in the classical limit.  
An interesting development in this direction was reported by Wheeler
\cite{Wheeler-SU3}, who wrote determinant formulas analogous to
\re{BetheVD} are obtained for the generalized model with $SU(3)$
symmetry.  

Finally, it is certainly worth to try to adjust the expression
\re{clasC123} for the non-compact rank-one sectors of the full
symmetry $PSU(2,2|4)$, such as $SL(2,\IR)$ and $SU(1,1)$, where the
integration contours should be placed along the real axis.

  \section*{Acknowledgments}
  
The author is obliged to S. Alexandrov, O. Foda, N. Gromov, C.
Kristjansen, A. Sever, D. Serban, F. Smirnov, P. Vieira and K. Zarembo
for many useful discussions and comments, and to O. Foda for a
critical reading of the manuscript.  Part of this work has been done
during the visit of the author at NORDITA in February 2012.

 \footnotesize
%
%

 \providecommand{\href}[2]{#2}\begingroup\raggedright\endgroup

 \end{document}